\documentclass[11pt]{article}
\usepackage{graphicx,caption2,subfigure}
\makeatletter
\newcommand{\singlespacing}{\let\CS=\@currsize\renewcommand{\baselinestretch}{1.0}\tiny\CS}
\newcommand{\doublespacing}{\let\CS=\@currsize\renewcommand{\baselinestretch}{1.5}\tiny\CS}

\begin{document}
\title {Rapidity and Pseudorapidity distributions of the Various Hadron-Species Produced in High Energy
 Nuclear Collisions : ~~~~~~~~~~~~~A Systematic Approach }
\author {Goutam
Sau$^1$\thanks{e-mail: sau$\_$goutam@yahoo.com}, A.
Bhattacharya$^2$\thanks{e-mail: pampa@phys.jdvu.ac.in} $\&$ S.
Bhattacharyya$^3$\thanks{e-mail: bsubrata@www.isical.ac.in
(Communicating Author).}\\
{\small $^1$ Beramara RamChandrapur High School,}\\
 {\small South 24-Pgs, 743609(WB), India.}\\
  {\small $^2$ Department of Physics,}\\
  {\small Jadavpur University, Kolkata- 700032, India. }\\
  {\small $^3$ Physics and Applied Mathematics Unit(PAMU),}\\
 {\small Indian Statistical Institute, Kolkata - 700108, India.}}
\date{}
\maketitle
\bigskip
\begin{abstract}
With the help of a phenomenological approach outlined in the text in
some detail, we have dealt here with the description of the plots on
rapidity and pseudorapidity spectra of some hadron-secondaries
produced in various nucleus-nucleus interactions at high energies.
The agreement between the measured data and the attempted fits are,
on the whole, modestly satisfactory excepting a very narrow central
region in the vicinity of y=$\eta$=0. At last, hints to how the
steps suggested in the main body of the text to proceed with the
description of the measured data given in the plots could lead
finally to a somewhat systematic methodology have also been made.
\end{abstract}
\newpage
\doublespacing
\section{Introduction and Background}
In a chain of our previous works we studied extensively the
properties of the rapidity (pseudorapidity) spectra of the various
secondary particles in some very high energy collisions. The left-
overs of the available latest data on rapidity (pseudorapidity)
spectra would here be dealt with for some specific secondaries
produced in some very high energy nuclear collisions. The secondary
species studied herein are mostly clearly identified particles. Our
objective here is quite clear. In the light of a Grand Combinational
Model (GCM) we would dwell here upon some aspects of the behaviour
of the rapidity (pseudorapidity) spectra of the identified
secondaries produced in high energy Au+Au and Pb+Pb interactions.
This would provide us an opportunity to check up the role of Grand
Combinational Model (GCM) in explaining some very recent and
interesting data on Au+Au and Pb+Pb reactions.
\par
Amidst the two previous works, this paper is more in line with our
second work\cite{Sau1} on understanding the nature of the rapidity
spectra for production of some heavy baryons than with the first
one. In this paper, an empirical energy-dependence of one of the
parameters was introduced and a systematic approach to the study was
built up. This work is just a follow-up of that particular
methodology. This work is essentially just an exhaustive study on
the rapidity-density or pseudorapidity-density of the identified
hadronic secondaries produced in some high energy nuclear
collisions. This apart, some clues to developing this procedure as a
systematic approach have also been highlighted.
\par
The paper is organized as follows. In the next section (Sec.2) we
give the basic outlook and the approach to be taken up for this
study. The following section (Sec.3) provides description of the
data analyses on Au+Au and Pb+Pb interactions mostly in the
graphical plots. The last section is reserved for summing up the
conclusions with some suggestive remarks to develop the applied
procedure into a somewhat complete systematic methodology.
\section{The Phenomenological Setting : Premises and the Pathway}
Following Faessler\cite{Faessler1}, Peitzmann\cite{Peitzmann1},
Schmidt and Schukraft\cite{Schmidt1} and finally Thom$\acute{e}$ et
al\cite{Thome1}, we \cite{De1,De2} had formulated in the past a
final working expression for rapidity distributions in proton-proton
collisions at ISR (Intersecting Storage Rings) ranges of
energy-values by the following three-parameter parametrization, viz,
\begin{equation}
\frac{1}{\sigma}\frac{d\sigma}{dy}=C_1(1+\exp\frac{y-y_0}{\Delta})^{-1}
\end{equation}
where $C_1$ is a normalization constant and $y_0$, $\Delta$ are two
parameters. The choice of the above form made by Thom$\acute{e}$ et
al\cite{Thome1} was intended to describe conveniently the central
plateau and the fall-off in the fragmentation region by means of the
parameters $y_0$ and $\Delta$ respectively. Besides, this was based
on the concept of both limiting fragmentation and the Feynman
Scaling hypothesis. For all five energies in PP collisions the value
of $\Delta$ was obtained to be $\sim$ 0.55 for pions\cite{De1} and
kaons\cite{De2}, $\sim$ 0.35 for protons/antiprotons\cite{De2}, and
$\sim$ 0.70 for $\Lambda$, $\Xi$, $\phi$, $\Sigma$ and $\Omega$. And
these values of $\Delta$ are generally assumed to remain the same in
the ISR ranges of energy. Still, for very high energies, and for
direct fragmentation processes which are quite feasible in very high
energy heavy nucleus-nucleus collisions, such parameter values do
change somewhat prominently, though in most cases with marginal high
energies, we have treated them as nearly constant.
\par Now, the fits for the rapidity (pseudorapidity)
 spectra for non-pion secondaries produced in the PP reactions at various energies are phenomenologically
 obtained by De and Bhattacharyya\cite{De2} through the making of
 suitable choices of $C_1$ and $y_0$. It is observed that for most of the secondaries the values of $y_0$
 do not remain exactly constant and show up some degree of species-dependence .
  However, for $\Lambda$, $\Xi$, $\Sigma$, $\Omega$ and $\phi$,  it gradually increases with energies and the energy-dependence
 of $y_0$ is empirically proposed to be expressed by the following relationship\cite{De1} :
\begin{equation}
y_0=k\ln\sqrt{s_{NN}}+0.8
\end{equation}
\par The nature of energy-dependence of $y_0$ is shown in the adjoining figure
(Fig.1). Admittedly, as k is assumed to vary very slowly with c. m.
energy, the parameter $y_0$ is not exactly linearly correlated to
$\ln \sqrt{s_{NN}}$, especially in the relatively low energy region.
And this is clearly manifested in Fig.1. This variation with energy
in k-values is introduced in order to accommodate and describe the
symmetry in the plots on the rapidity spectra around mid-rapidity.
This is just phenomenologically observed by us, though we cannot
readily provide any physical justification for such perception
and/or observation. And the energy-dependence of $y_0$ is studied
here just for gaining insights in their nature and for purposes of
extrapolation to the various higher energies (in the centre of mass
frame, $\sqrt{s_{NN}}$) for several nucleon-nucleon, nucleon-nucleus
and nucleus-nucleus collisions. The specific energy (in the c.m.
system,
 $\sqrt{s_{NN}}$) for every nucleon-nucleus or nucleus-nucleus collision is first worked out by
 converting the laboratory energy value(s) in the required c.m. frame energy value(s).
 Thereafter the value of $y_0$ to be used for computations of inclusive cross-sections of nucleon-nucleon collisions
 at particular energies of interactions is extracted from Eq. (2) for corresponding obtained energies.
 This procedural step is followed for calculating the rapidity (pseudorapidity)-spectra for not only the pions
 produced in nucleon-nucleus and nucleus-nucleus collisions\cite{De1}. However, for the studies on the rapidity-spectra
 of the non-pion secondaries produced in the same reactions one does always neither have the opportunity to take recourse
 to such a systematic step, nor could they actually resort to this rigorous
 procedure, due to the lack of necessary and systematic data on
 them.
\par Our next step is to explore the nature of $f(y)$ which is
envisaged to be given generally by a polynomial form noted
below :
\begin{equation}
f(y) = \alpha + \beta y + \gamma y^2,
\end{equation}
where $\alpha$, $\beta$ and $\gamma$ are the coefficients to be
chosen separately for each AB collisions (and also for AA collisions
when the projectile and the target are same). Besides, some other
points are to be made here. The suggested choice of form in
expression (3) is not altogether fortuitous. In fact, we got the
clue from one of the previous work by one of the authors
(SB)\cite{Bhattacharyya1} here pertaining to the studies on the
behavior of the EMC effect related to the lepto-nuclear collisions.
In the recent past Hwa et al\cite{Hwa1} also made use of this sort
of relation in a somewhat different context. Now let us revert to
our original discussion and to the final working formula for
$\frac{dN}{dy}$ in various AB (or AA) collisions given by the
following relation :
\begin{equation}
\frac{dN}{dy}|_{AB \rightarrow QX} = C_2(AB)^{\alpha + \beta y + \gamma y^2}\frac{dN}{dy}|_{PP \rightarrow QX}
 = C_3(AB)^{\beta y + \gamma y^2}(1 + \exp \frac{y-y_0}{\Delta})^{-1},
\end{equation}
where $C_2$ is the normalization constant and $C_3$=$C_2(AB)^\alpha$
is another constant as $\alpha$ is also a constant for a specific
collision at a specific energy. The parameter values for different
nucleus-nucleus collisions are given in the Tables (Table2 - Table
11).
\par
However, it is to be noted that the relationship between rapidity
and pseudorapidity is given by the following standard relation
\begin{equation}
\frac{dN}{d\eta dp_T^2}=\sqrt{1-\frac{m^2}{m_T^2 cosh^2y}}\frac{dN}{dy dp_T^2}
\end{equation}
with the following properties :
\par
(a) In the region y$>>$0, $\frac{dN}{d\eta}\approx \frac{dN}{dy}$
\par
(b) But, in the region y$\rightarrow$0, there ia a small depression
of the $\frac{dN}{d\eta}$ distribution relative to $\frac{dN}{dy}$
due to the above transformation. In experiments at high energies
where $\frac{dN}{dy}$ has a plateau shape, this transformation gives
a small dip in $\frac{dN}{d\eta}$ around $\eta \approx$0.
\par
(c) In the c.m. frame, the peak of the distribution is located
around y$\approx \eta \approx$0, and the peak value of
$\frac{dN}{d\eta}$ is smaller than the peak value of
$\frac{dN}{dy}$; And this Diminutive Fraction Factor (DFF) is given
by
\begin{equation}
DFF \approx [1-\frac{m^2}{<m_T^2>}]^{\frac{1}{2}}
\end{equation}
\section{Depicting the Results Obtained}
\subsection{A Few Pointed Steps}
The procedural steps for arriving at the results could be summed up
as follows :
\par (i) We assume that the inclusive
cross section (I.C.) of any particle in a nucleus-nucleus (AB)
collision can be obtained from the production of the same in
nucleon-nucleon collisions by multiplying the inclusive
cross-section (I.C.) by a product of the atomic numbers of each of
the colliding nuclei raised to a particular function, which is
initially unspecified\cite{Sau2}.
\par (ii) Secondly, we accept the property of factorization\cite{Sau1} of that particular function
which helps us to perform the integral over $p_T$ in a relatively
simpler manner.
\par (iii) Thirdly, we assume
a particular 3-parameter form for the pp cross section with the
parameters $C_1$, $y_0$ and $\Delta$.
\par (iv) Finally,
we accept the ansatz that the function f(y) can be modeled by a
quadratic function with the parameters $\alpha$, $\beta$ and
$\gamma$.
\subsection{Final Results Delivered}
The results are shown here by the graphical plots with the
accompanying tables for the parameter values. Here we draw the
rapidity-density of pion($\pi$), kaon($K$), proton-antiproton($N$),
$\phi$, $\Omega$, $\Sigma$, $\Lambda$, $\Xi$ for symmetric Pb+Pb and
Au+Au collisions and pseudorapidity-density of charged-particle
(mainly $\pi^+$) for symmetric Au+Au collisions at several energies
which have been appropriately labeled at the top right corner. In
this context some comments are in order. Though the figures
represents the case for production of pion($\pi$), kaon($K$),
proton-antiproton($N$), $\phi$, $\Omega$, $\Sigma$, $\Lambda$,
$\Xi$, we do not anticipate and/or expect any strong
charge-dependence of the results. Besides, the solid curves in all
cases-almost without any exception-demonstrate our GCM-based
results. Secondly, the data on rapidity(pseudorapidty)-spectra for
some high-energy collisions are, at times, available for both
positive and negative y($\eta$)-values. This gives rise to a problem
in our method. It is evident here in this work that we are concerned
with only symmetric collisions wherein the colliding nuclei must be
identical. But in our expression (4) the coefficient $\beta$
multiplies a term which is proportional to y and so is not symmetric
under y$\rightarrow$(-y). In order to overcome this difficulty we
would introduce here $\beta$=0 for all the graphical plots (except
Fig.12). These plots are represented by Fig.2 to Fig.11 for $\pi, K,
\phi, N, \Sigma, \Xi, \Lambda, \Omega$ in Pb+Pb and Au+Au collision
under different conditions. The parameter values in this particular
case are presented in tables (Table 2 - Table 10). The graphical
plots shown in fig.2 and Fig.3 (for $\beta$=0) are for production of
$\pi^-, K^-, K^+, \phi$ in Pb+Pb collisions at 20A GeV, 30A GeV, 40A
GeV, 80A GeV respectively. The diagrams shown in Fig.4 represent the
production of $\pi, N, K, \Sigma, \Lambda, \Xi, \Omega$ in Au+Au
interaction at $\sqrt{s_(NN)}$=7 GeV (for $\beta$=0). And the plots
depicted for pseudorapidity-spectra in Fig.5 to Fig.11 are based on
the production of charged particle (we consider only $\pi^+$) in
Au+Au collision for different centrality bins at 19.6 GeV, 62.4 GeV,
130 GeV and 200 GeV respectively (for $\beta$=0). The plots shown in
Fig.12 are for the production of charged particle (mainly $\pi^+$)
for four different energies i.e., 19.6 GeV, 62.4 GeV, 130 GeV, 200
GeV respectively and the parameter values are shown in Table 11.
Here we would mention that the data are for 19.6 GeV, 130 GeV, 200
Gev for PHOBOS Collaboration and that of 62.4 GeV for STAR
Collaboration as shown on the top right corner of the figure.
Finally, the diagram in Fig.13 represents the variation of $\beta$
and $\gamma$ with the energy values and we draw a mean curve in this
particular diagram.
\section{Concluding Remarks and Some Comments}
On an overall basis, our model-based results are in fair agreement
with the most of the data-sets, excepting y$\approx$0 or
$\eta$$\approx$0 region, wherein the data shows flat-plateau
structures in almost all the diagrams exhibiting data on both
positive and negative rapidities or pseudorapidities. The degree of
disagreement in the vicinity of $\eta$$\approx$0 region is evidently
much stronger for the plots on pseudorapidity-density versus
pseudorapidity plots. These discrepancies might probably be ascribed
to our simplistic assumption of y=$\eta$. Had we been able to
compute the diminutive fraction factor (DFF) as given by expression
(6), we would have been capable of giving the pseudorapidity-figures
much better looks. And this computation is not possible because of
the fact that the rapidity-data-sets do not generally offer even the
slightest hints on the $p_T$-ranges of the secondaries under
observations and/or measurements.
\par
The last figure of this paper carries some special physical
significance which we now explain below. This is, by essence,
undoubtedly a purely phenomenological model with no or very little
predictive capacity. The energy-dependences studied in Fig. 13 for
some of the involved parameters, $\beta$ and $\gamma$, could provide
us some insights into what could be the possible values of $\beta$
and $\gamma$ at some higher/lower/intermediate values of the c.m.
energies of the interactions for any specific secondary. This could
help, we believe, to reduce the elements/components of phenomenology
and introduce some degree, however low, of predictivity of values of
$\beta$ and $\gamma$ by necessary intrapolation or extrapolation, as
the case may be, for any specific secondary produced in the same
nuclear interactions. Thus, if sufficient and reliable data at, at
least, six to seven c.m. energies at reasonable intervals are
available allowing the scopes for studying the nature of c.m.
energy-dependence of these parameters, the present procedure could
be nurtured to a better and more competent methodical approach.

\newpage
{\singlespacing{
 \begin{table}
\begin{center}
\begin{small}
\caption{Variation of $y_0$ with Energy.[Reference Fig. No.1]}
\begin{tabular}{|c|c|c|}\hline
 $Energy(\sqrt{s_{NN}})(GeV)$ & $Constant (k)$ & $y_0$ \\
  \hline
  $6.3 (20 AGeV)$ & $ 2.76$ & $5.894$ \\
 \hline
  $7$ & $ 2.65$ & $5.951$ \\
  \hline
  $7.6 (30 AGeV)$ & $ 2.54 $ & $6.006$ \\
  \hline
  $8.7 (40 AGeV)$ & $ 2.40 $ & $6.085$ \\
  \hline
  $12.3 (80 AGeV)$ & $ 2.16 $ & $6.276$ \\
 \hline
  $19.6$ & $ 1.92$ & $6.517$ \\
 \hline
  $62.4$ & $1.54$ & $7.153$\\
 \hline
  $130$ & $1.39$ & $7.556$\\
 \hline
 $200$ & $1.32$ & $7.794$\\
 \hline
 \end{tabular}
\end{small}
\end{center}

\begin{center}
\begin{small}
\caption{Values of different parameters for production of identified
hadrons in central Pb+Pb collisions at $E_{beam}$ = 20 AGeV (for
$\beta$=0) for both +ve and -ve rapidities.[Reference Fig. No.2(a)
\& 3(a)]}
\begin{tabular}{|c|c|c|c|}\hline
 $Production$ & $C_3$ & $\gamma$ & $\frac{\chi^2}{ndf}$\\
 \hline
 $\pi^-$ & $84.941 \pm0.2108 $  & $-0.044 \pm0.0003$ & $21.509/22 $ \\
 \hline
  $K^+$ & $20.574\pm0.1556 $ & $-0.057 \pm0.0008 $& $4.119/05 $ \\
 \hline
 $K^-$ & $05.507 \pm0.0447 $  & $-0.091 \pm0.0015$ & $7.753/14 $ \\
 \hline
  $\phi$ & $01.325\pm0.0406 $ & $-0.139 \pm0.0021 $& $4.423/07$ \\
 \hline
\end{tabular}
\end{small}
\end{center}

\begin{center}
\begin{small}
\caption{Values of different parameters for production of identified
hadrons in central Pb+Pb collisions at $E_{beam}$ = 30 AGeV (for
$\beta$=0) for both +ve and -ve rapidities.[Reference Fig. No.2(b)
\& 3(b)]}
\begin{tabular}{|c|c|c|c|}\hline
 $Production$ & $C_3$ & $\gamma$ & $\frac{\chi^2}{ndf}$\\
 \hline
 $\pi^-$ & $97.912 \pm0.2754 $  & $-0.037 \pm0.0003$ & $10.898/13 $ \\
 \hline
  $K^+$ & $22.849\pm0.1128 $ & $-0.049 \pm0.0005 $& $5.824/08 $ \\
 \hline
 $K^-$ & $07.547\pm0.0705 $  & $-0.075 \pm0.0011$ & $4.316/10 $ \\
 \hline
  $\phi$ & $01.384\pm0.0158 $ & $-0.095 \pm0.0008 $& $4.264/06$ \\
 \hline
\end{tabular}
\end{small}
\end{center}

\begin{center}
\begin{small}
\caption{Values of different parameters for production of identified
hadrons in central Pb+Pb collisions at $E_{beam}$ = 40 AGeV (for
$\beta$=0) for both +ve and -ve rapidities.[Reference Fig. No.2(c)
\& 3(c)]}
\begin{tabular}{|c|c|c|c|}\hline
 $Production$ & $C_3$ & $\gamma$ & $\frac{\chi^2}{ndf}$\\
 \hline
 $\pi^-$ & $111.998\pm0.3811 $  & $-0.035 \pm0.0003$ & $8.926/10 $ \\
 \hline
  $K^+$ & $023.363\pm0.1275 $ & $-0.039 \pm0.0005 $& $4.621/06 $ \\
 \hline
 $K^-$ & $010.605\pm0.0755 $  & $-0.073 \pm0.0003$ & $3.394/09 $ \\
 \hline
  $\phi$ & $001.185\pm0.0147 $ & $-0.063 \pm0.0006 $& $4.206/09$ \\
 \hline
\end{tabular}
\end{small}
\end{center}
\end{table}

\begin{table}
\begin{center}
\begin{small}
\caption{Values of different parameters for production of identified
hadrons in central Pb+Pb collisions at $E_{beam}$ = 80 AGeV (for
$\beta$=0) for both +ve and -ve rapidities.[Reference Fig. No.2(d)
\& 3(d)]}
\begin{tabular}{|c|c|c|c|}\hline
 $Production$ & $C_3$ & $\gamma$ & $\frac{\chi^2}{ndf}$\\
 \hline
 $\pi^-$ & $147.686\pm0.4383 $  & $-0.027 \pm0.0002$ & $8.659/11 $ \\
 \hline
  $K^+$ & $026.426\pm0.1038 $ & $-0.030 \pm0.0004 $& $5.455/08 $ \\
 \hline
 $K^-$ & $012.790\pm0.0763 $  & $-0.041 \pm0.0007$ & $9.002/12 $ \\
 \hline
  $\phi$ & $001.762\pm0.0044 $ & $-0.044 \pm0.0002 $& $0.543/06$ \\
 \hline
\end{tabular}
\end{small}
\end{center}

\begin{center}
\begin{small}
\caption{Values of different parameters for production of identified
hadrons in central Au+Au collisions at $\sqrt{s_{NN}}$ = 7 GeV (for
$\beta$=0) for both +ve and -ve rapidities.[Reference Fig. No.4]}
\begin{tabular}{|c|c|c|c|}\hline
 $Production$ & $C_3$ & $\gamma$ & $\frac{\chi^2}{ndf}$\\
 \hline
 $\pi$ & $299.742\pm0.4296 $  & $-0.026 \pm0.0001$ & $8.014/07 $ \\
 \hline
  $N$ & $136.702\pm0.2569 $ & $-0.048 \pm0.0005 $& $5.460/04 $ \\
 \hline
 $K$ & $011.734\pm0.0248 $  & $-0.030 \pm0.0001$ & $0.621/03 $ \\
 \hline
  $\Sigma, \Lambda$ & $015.886\pm0.0055 $ & $-0.056 \pm0.0004 $& $4.750/05$ \\
 \hline
 $\Omega$ & $000.017\pm0.0004 $ & $-0.079 \pm0.0048 $& $8.754/10$ \\
 \hline
 $\Xi$ & $000.628\pm0.0037 $ & $-0.057 \pm0.0010 $& $9.502/14$ \\
 \hline
\end{tabular}
\end{small}
\end{center}

\begin{center}
\begin{small}
\caption{Values of different parameters for production of
charged-particle ($\pi^+$) in Au+Au collisions at $\sqrt{s_{NN}}$ =
19.6 GeV (for $\beta$=0) for both +ve and -ve
pseudo-rapidities.[Reference Fig. No.5]}
\begin{tabular}{|c|c|c|c|}\hline
 $Centrality$ & $C_3$ & $\gamma$ & $\frac{\chi^2}{ndf}$\\
 \hline
 $03\%-06\%$ & $363.098\pm1.764 $  & $-0.0157\pm0.00009$ & $17.359/12 $ \\
 \hline
  $06\%-10\%$ & $328.094\pm1.195 $ & $-0.0148 \pm0.00007 $& $12.873/19 $ \\
 \hline
 $10\%-15\%$ & $266.638\pm0.967 $  & $-0.0137 \pm0.00007$ & $21.209/18 $ \\
 \hline
  $15\%-20\%$ & $207.392\pm1.130 $ & $-0.0125 \pm0.00008 $& $9.160/10$ \\
 \hline
 $20\%-25\%$ & $180.106\pm0.656 $ & $-0.0123 \pm0.00007 $& $15.549/16$ \\
 \hline
 $25\%-30\%$ & $150.634\pm0.456 $ & $-0.0117 \pm0.00006 $& $14.747/12$ \\
 \hline
 $30\%-35\%$ & $115.939\pm0.436 $ & $-0.0110 \pm0.00007 $& $16.130/15$ \\
 \hline
 $35\%-40\%$ & $087.982\pm0.405 $ & $-0.0098 \pm0.00008 $& $11.447/13$ \\
 \hline
 $40\%-45\%$ & $071.793\pm0.249 $ & $-0.0093 \pm0.00008$& $16.840/10$ \\
 \hline
\end{tabular}
\end{small}
\end{center}

\begin{center}
\begin{small}
\caption{Values of different parameters for production of
charged-particle ($\pi^+$) in Au+Au collisions at $\sqrt{s_{NN}}$ =
62.4 GeV (for $\beta$=0) for both +ve and -ve
pseudo-rapidities.[Reference Fig. No.6 \& 7]}
\begin{tabular}{|c|c|c|c|}\hline
 $Centrality$ & $C_3$ & $\gamma$ & $\frac{\chi^2}{ndf}$\\
 \hline
 $00\%-03\%$ & $858.977\pm2.349 $  & $-0.0131\pm0.00002$ & $3.527/04 $ \\
 \hline
 $03\%-06\%$ & $805.167\pm1.252 $  & $-0.0135\pm0.00008$ & $9.541/12 $ \\
 \hline
  $06\%-10\%$ & $690.250\pm4.553 $ & $-0.0126 \pm0.00006 $& $15.843/13 $ \\
 \hline
 $10\%-15\%$ & $505.765\pm2.056 $  & $-0.0111 \pm0.00004$ & $15.882/15 $ \\
 \hline
  $15\%-20\%$ & $397.679\pm1.516 $ & $-0.0101 \pm0.00004 $& $11.527/12$ \\
 \hline
 $20\%-25\%$ & $288.758\pm1.128 $ & $-0.0086 \pm0.00005 $& $9.042/10$ \\
 \hline
 $25\%-30\%$ & $223.869\pm1.019 $ & $-0.0081 \pm0.00004 $& $8.544/08$ \\
 \hline
 $30\%-35\%$ & $179.875\pm0.496 $ & $-0.0076 \pm0.00004 $& $10.873/11$ \\
 \hline
 $35\%-40\%$ & $126.863\pm0.661 $ & $-0.0066 \pm0.00005 $& $10.169/10$ \\
 \hline
 $40\%-45\%$ & $112.144\pm0.404 $ & $-0.0068 \pm0.00005$& $6.154/05$ \\
 \hline
 $45\%-50\%$ & $091.802\pm0.290 $ & $-0.0077 \pm0.00004$& $11.346/09$ \\
 \hline
\end{tabular}
\end{small}
\end{center}
\end{table}

\begin{table}
\begin{center}
\begin{small}
\caption{Values of different parameters for production of
charged-particle ($\pi^+$) in Au+Au collisions at $\sqrt{s_{NN}}$ =
130 GeV (for $\beta$=0) for both +ve and -ve
pseudo-rapidities.[Reference Fig. No.8 \& 9]}
\begin{tabular}{|c|c|c|c|}\hline
 $Centrality$ & $C_3$ & $\gamma$ & $\frac{\chi^2}{ndf}$\\
 \hline
 $00\%-03\%$ & $1017.26\pm4.767 $  & $-0.0106\pm0.00006$ & $8.489/08 $ \\
 \hline
 $03\%-06\%$ & $859.123\pm4.007 $  & $-0.0096\pm0.00005$ & $11.504/11 $ \\
 \hline
  $06\%-10\%$ & $771.332\pm4.214 $ & $-0.0094 \pm0.00005 $& $11.448/11 $ \\
 \hline
 $10\%-15\%$ & $553.134\pm1.895 $  & $-0.0083 \pm0.00002$ & $11.348/07 $ \\
 \hline
  $15\%-20\%$ & $490.218\pm2.056 $ & $-0.0084 \pm0.00004 $& $13.212/15$ \\
 \hline
 $20\%-25\%$ & $384.377\pm1.812 $ & $-0.0079 \pm0.00004 $& $9.556/16$ \\
 \hline
 $25\%-30\%$ & $342.650\pm1.736 $ & $-0.0081 \pm0.00004 $& $14.193/13$ \\
 \hline
 $30\%-35\%$ & $264.987\pm1.080 $ & $-0.0077 \pm0.00004 $& $8.711/13$ \\
 \hline
 $35\%-40\%$ & $183.701\pm0.757 $ & $-0.0061 \pm0.00005 $& $3.271/06$ \\
 \hline
 $40\%-45\%$ & $165.105\pm0.243 $ & $-0.0068 \pm0.00001$& $11.067/08$ \\
 \hline
 $45\%-50\%$ & $115.942\pm0.417 $ & $-0.0065 \pm0.00003$& $8.366/13$ \\
 \hline
\end{tabular}
\end{small}
\end{center}

\begin{center}
\begin{small}
\caption{Values of different parameters for production of
charged-particle ($\pi^+$) in Au+Au collisions at $\sqrt{s_{NN}}$ =
200 GeV (for $\beta$=0) for both +ve and -ve
pseudo-rapidities.[Reference Fig. No.10 \& 11]}
\begin{tabular}{|c|c|c|c|}\hline
 $Centrality$ & $C_3$ & $\gamma$ & $\frac{\chi^2}{ndf}$\\
 \hline
 $00\%-03\%$ & $1305.84\pm6.334 $  & $-0.0095\pm0.00004$ & $8.374/07 $ \\
 \hline
 $03\%-06\%$ & $1217.02\pm3.923 $  & $-0.0092\pm0.00002$ & $2.677/07 $ \\
 \hline
  $06\%-10\%$ & $1085.85\pm5.787 $ & $-0.0089 \pm0.00003 $& $8.494/08 $ \\
 \hline
 $10\%-15\%$ & $786.155\pm5.756 $  & $-0.0083 \pm0.00004$ & $10.695/07 $ \\
 \hline
  $15\%-20\%$ & $671.544\pm0.816 $ & $-0.0080 \pm0.00001 $& $13.220/08$ \\
 \hline
 $20\%-25\%$ & $494.043\pm4.083 $ & $-0.0076 \pm0.00005 $& $8.479/08$ \\
 \hline
 $25\%-30\%$ & $371.719\pm2.104 $ & $-0.0069 \pm0.00004 $& $7.826/06$ \\
 \hline
 $30\%-35\%$ & $298.854\pm2.012 $ & $-0.0069 \pm0.00005 $& $10.232/07$ \\
 \hline
 $35\%-40\%$ & $298.430\pm1.067 $ & $-0.0077 \pm0.00001 $& $19.517/16$ \\
 \hline
 $40\%-45\%$ & $142.160\pm0.609 $ & $-0.0060 \pm0.00003$& $11.112/10$ \\
 \hline
 $45\%-50\%$ & $203.956\pm0.874 $ & $-0.0063 \pm0.00003$& $13.112/08$ \\
 \hline
\end{tabular}
\end{small}
\end{center}

\begin{center}
\begin{small}
\caption{Values of different parameters for production of
charged-particle ($\pi^+$) in Au+Au collisions at four different
energies for +ve pseudo-rapiditie only.[Reference Fig. No.12]}
\begin{tabular}{|c|c|c|c|c|}\hline
 $Energy (GeV)$ & $C_3$ & $\beta$ & $\gamma$ & $\frac{\chi^2}{ndf}$\\
 \hline
 $19.6$ & $361.238\pm3.900 $ & $0.015\pm0.00032 $  & $-0.020\pm0.00009$ & $8.596/25 $ \\
 \hline
 $62.4$ & $778.667\pm8.000 $ & $0.018\pm0.00028 $   & $-0.018\pm0.00008$ & $3.847/04 $ \\
 \hline
  $130$ & $511.297\pm7.890 $  & $0.027\pm0.00054 $  & $-0.012 \pm0.00015$ & $14.445/25 $ \\
 \hline
 $200$ & $557.934\pm7.087 $ & $0.028\pm0.00044 $  & $-0.011 \pm0.00011 $& $9.578/26 $ \\
 \hline
 \end{tabular}
\end{small}
\end{center}
\end{table}

\newpage
\begin{figure}
\centering
\includegraphics[width=2.5in]{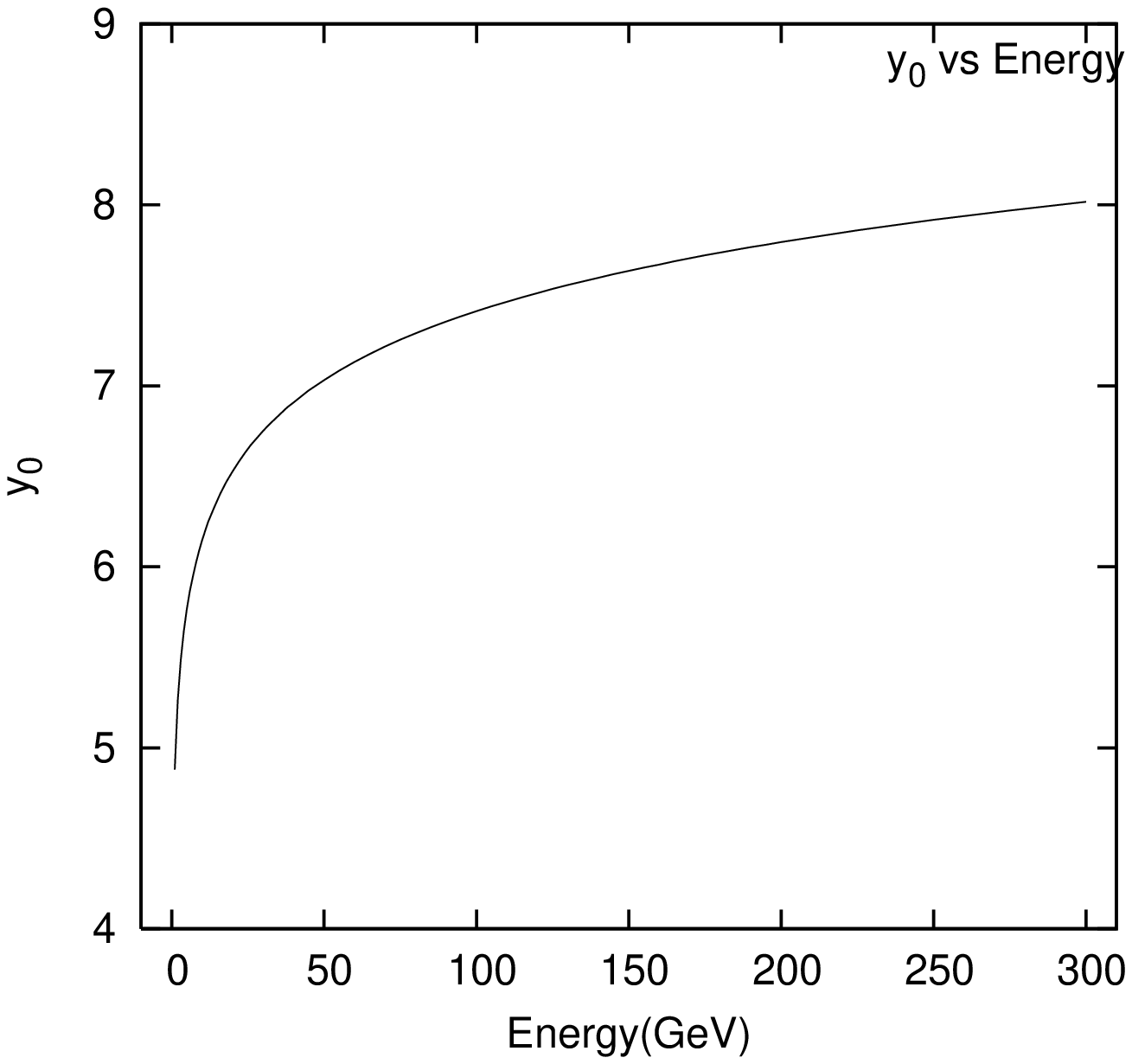}
\caption{Variation of $y_0$ in equation (2) with increasing energy.[Parameter values are shown in Table 1.]}

\subfigure[]{
\begin{minipage}{.5\textwidth}
\centering
\includegraphics[width=2.5in]{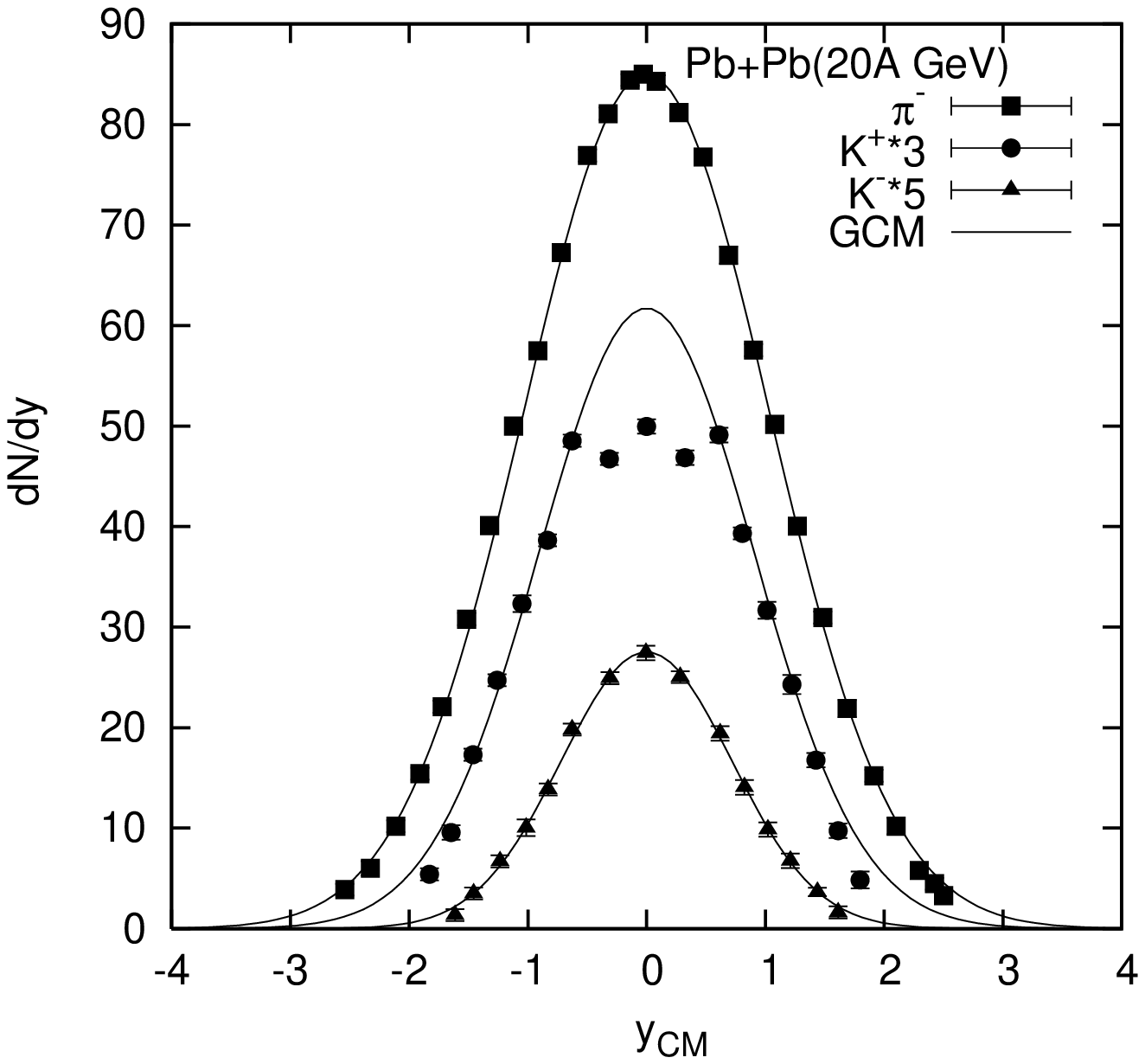}
\setcaptionwidth{2.6in}
\end{minipage}}%
\subfigure[]{
\begin{minipage}{0.5\textwidth}
\centering
 \includegraphics[width=2.5in]{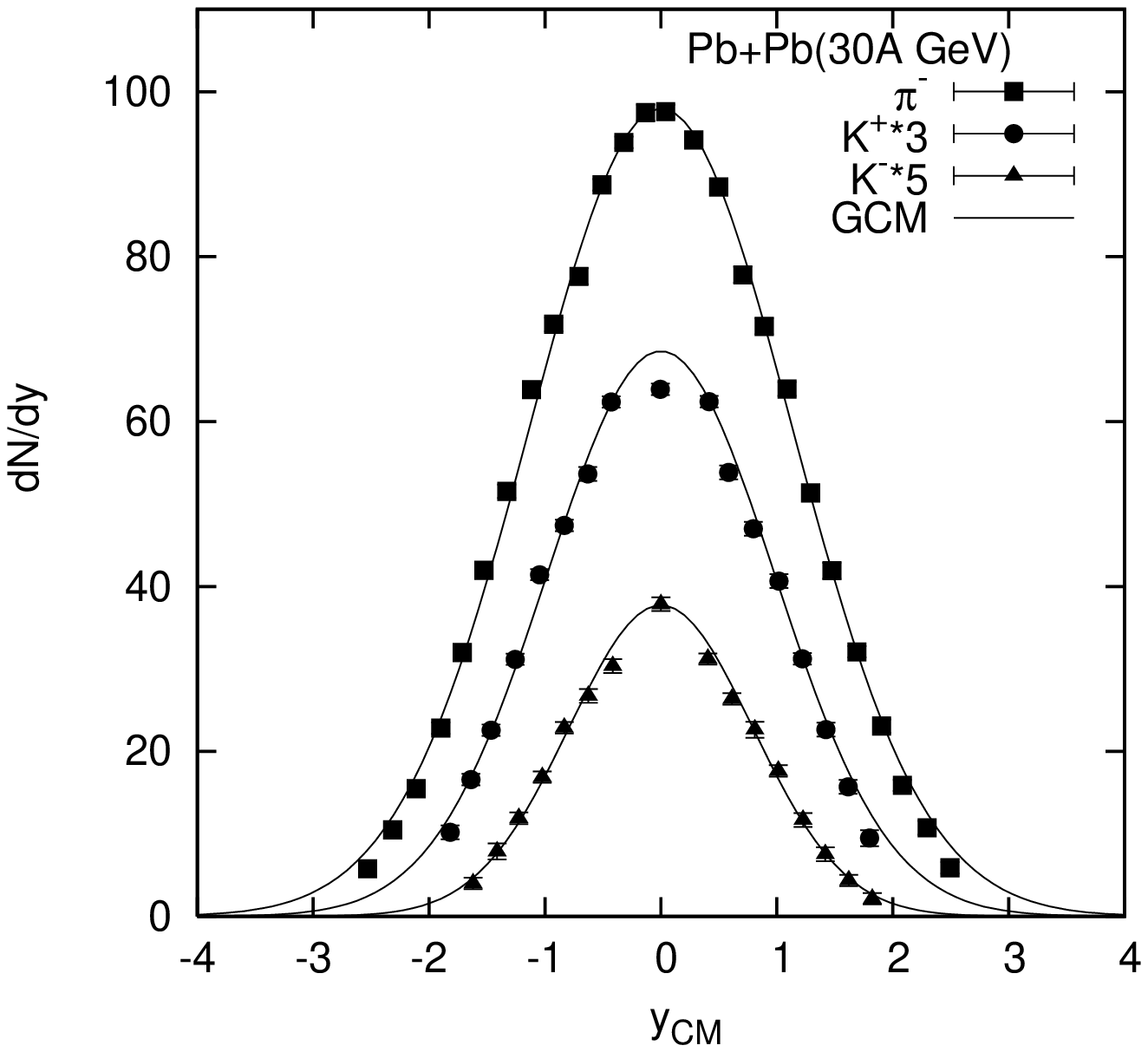}
 \end{minipage}}%
\vspace{0.01in} \subfigure[]{
\begin{minipage}{0.5\textwidth}
\centering
\includegraphics[width=2.5in]{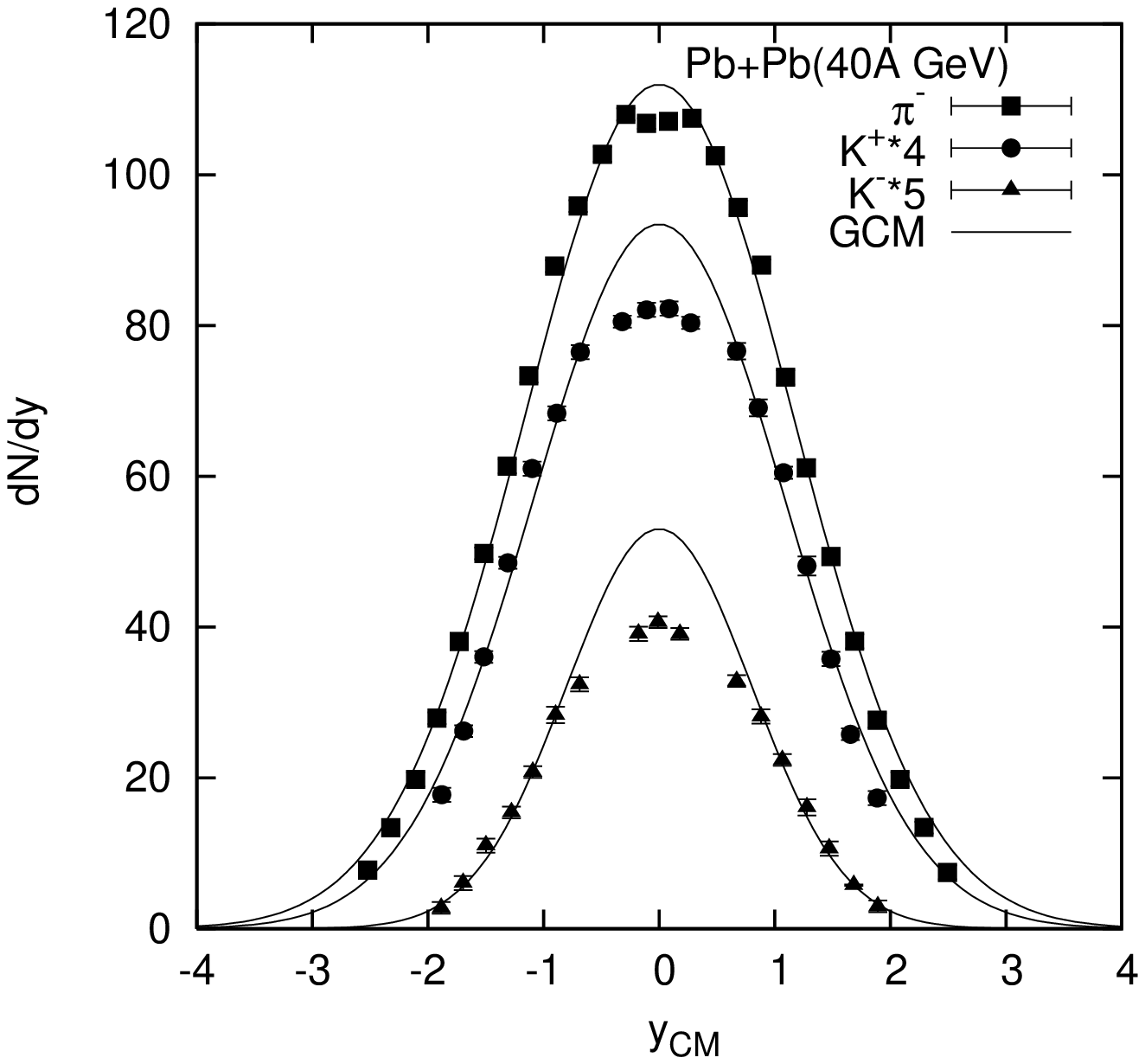}
\end{minipage}}%
\subfigure[]{
\begin{minipage}{.5\textwidth}
\centering
 \includegraphics[width=2.5in]{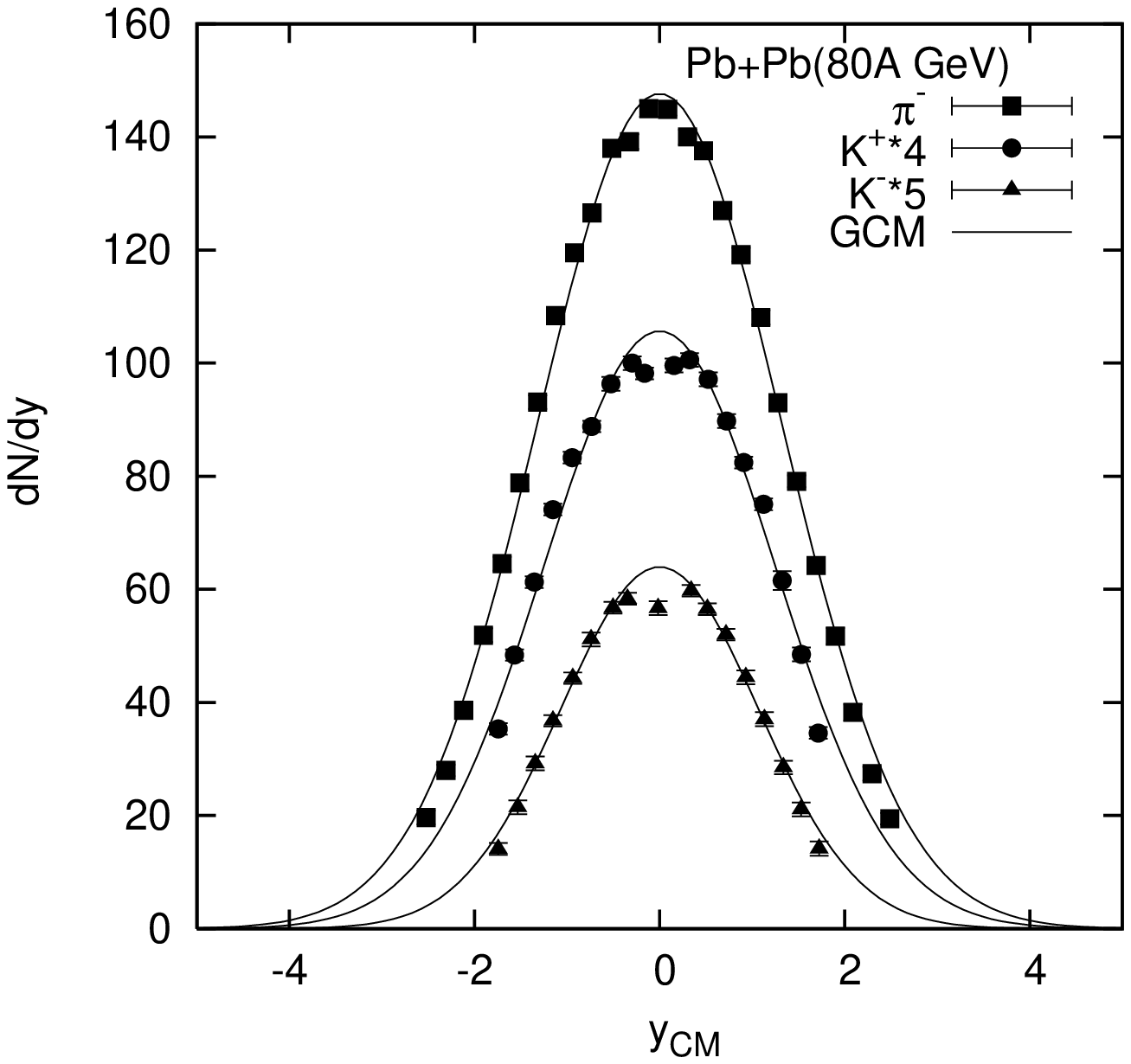}
 \end{minipage}}%
\caption{Rapidity distributions of identified hadrons in central Pb+Pb collisions at $E_{beam}$ = 20, 30, 40, 80
AGeV for $\beta$=0. The symbols are the experimental data and the data points are taken from {\cite{Xue1}} and the parameter
values are taken from Table 2-Table 5. The solid curve provide the GCM-based results.}
\end{figure}

\begin{figure}
\subfigure[]{
\begin{minipage}{.5\textwidth}
\centering
\includegraphics[width=2.5in]{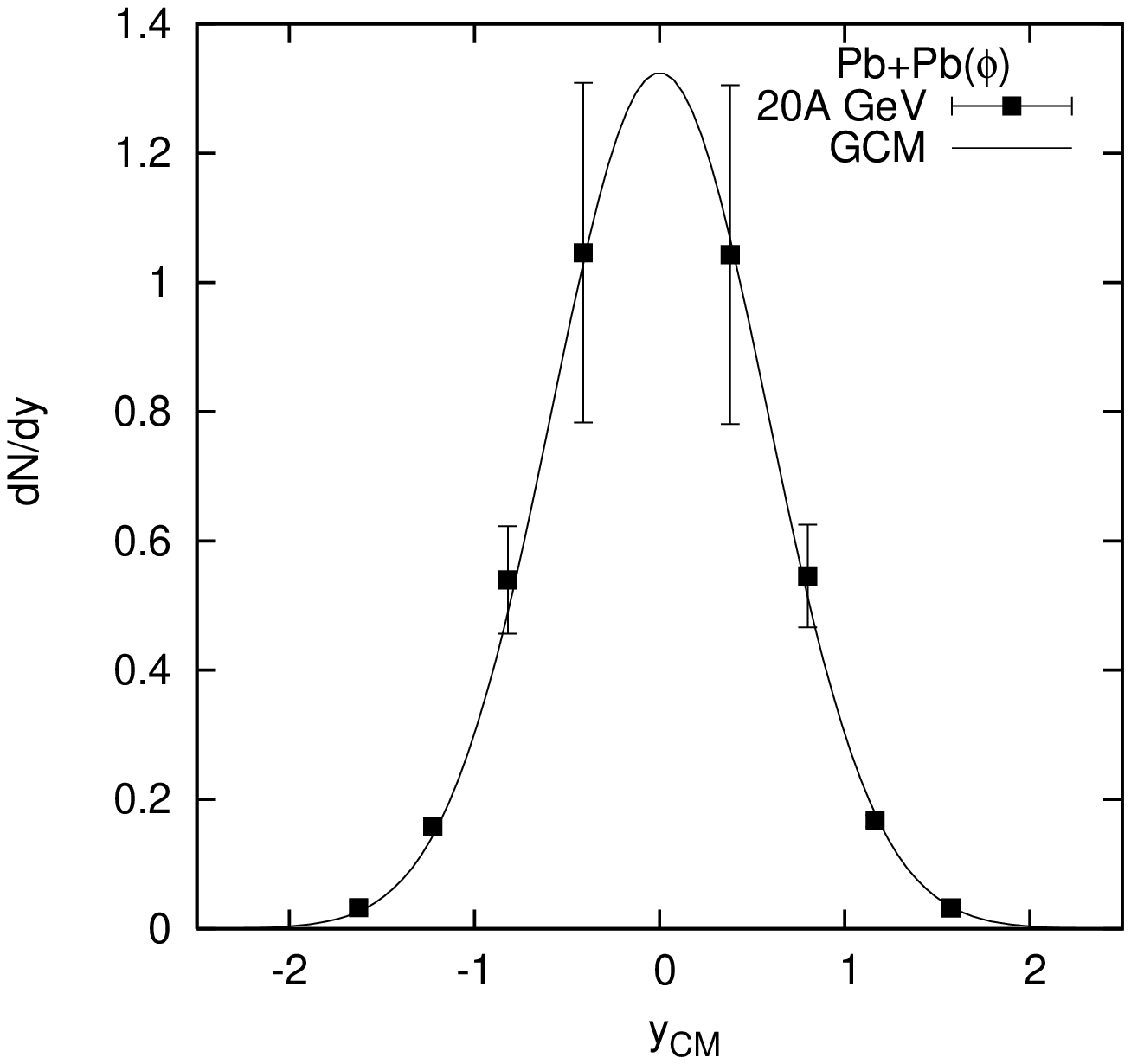}
\setcaptionwidth{2.6in}
\end{minipage}}%
\subfigure[]{
\begin{minipage}{0.5\textwidth}
\centering
 \includegraphics[width=2.5in]{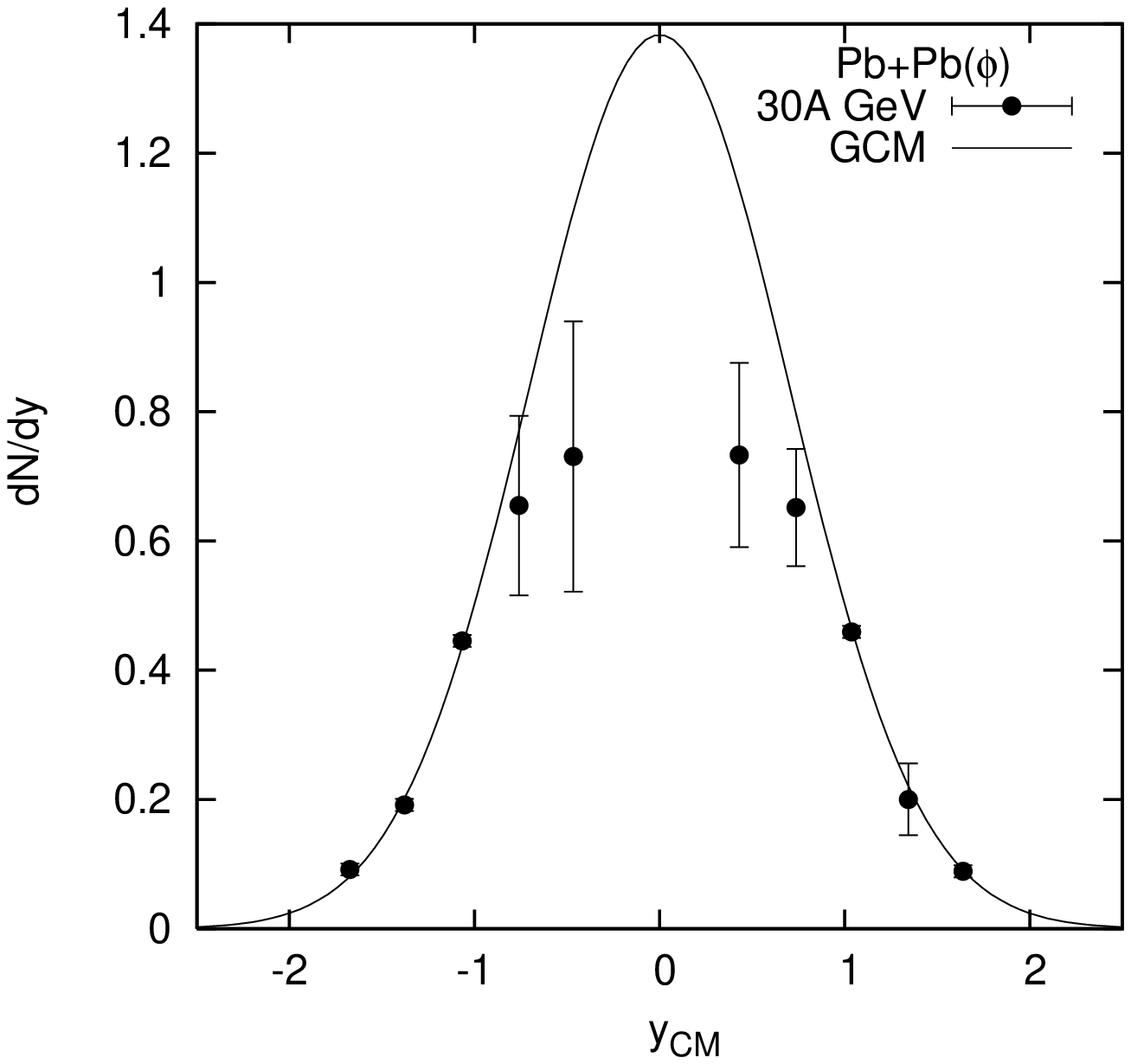}
 \end{minipage}}%
\vspace{0.01in} \subfigure[]{
\begin{minipage}{0.5\textwidth}
\centering
\includegraphics[width=2.5in]{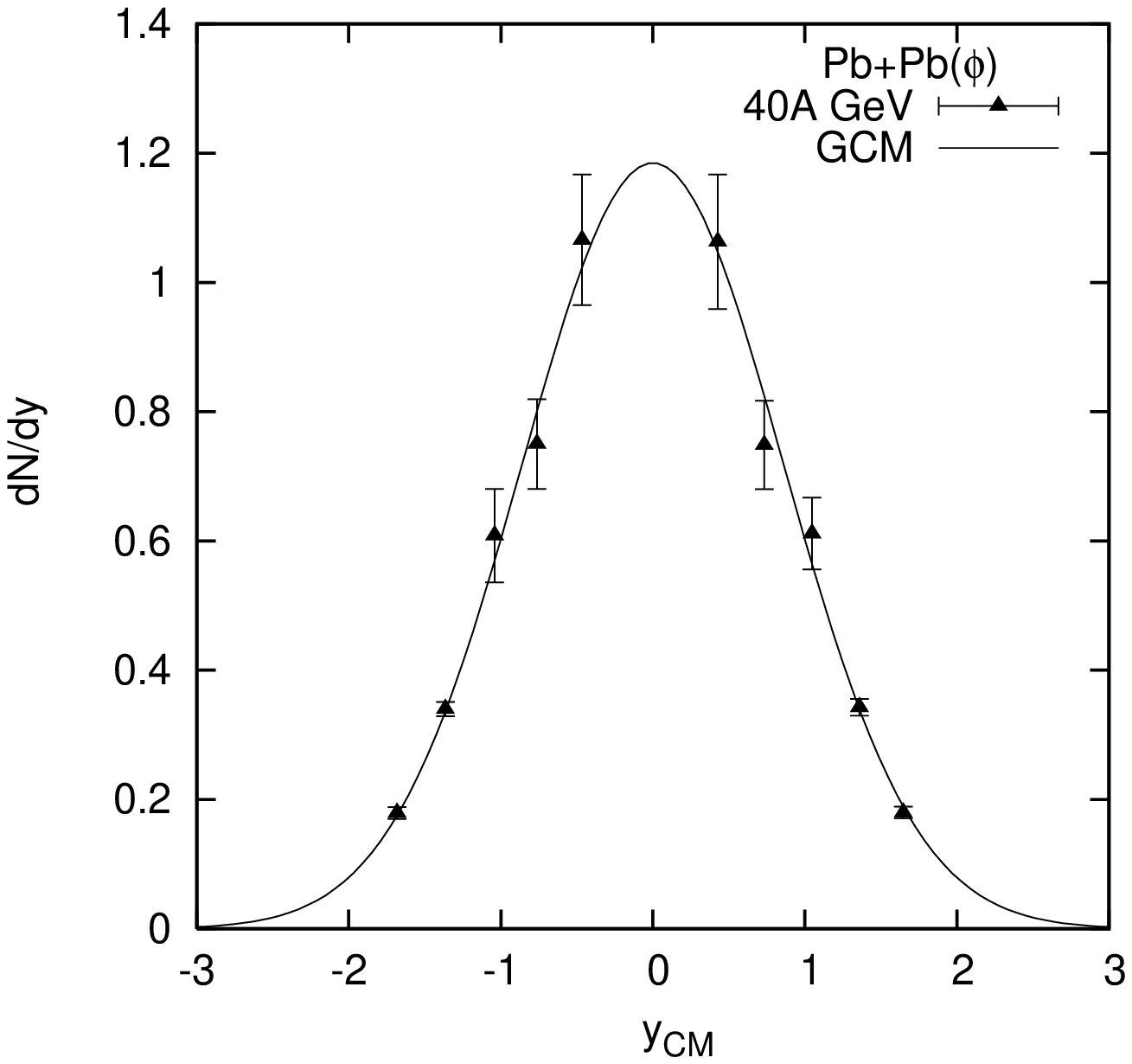}
\end{minipage}}%
\subfigure[]{
\begin{minipage}{.5\textwidth}
\centering
 \includegraphics[width=2.5in]{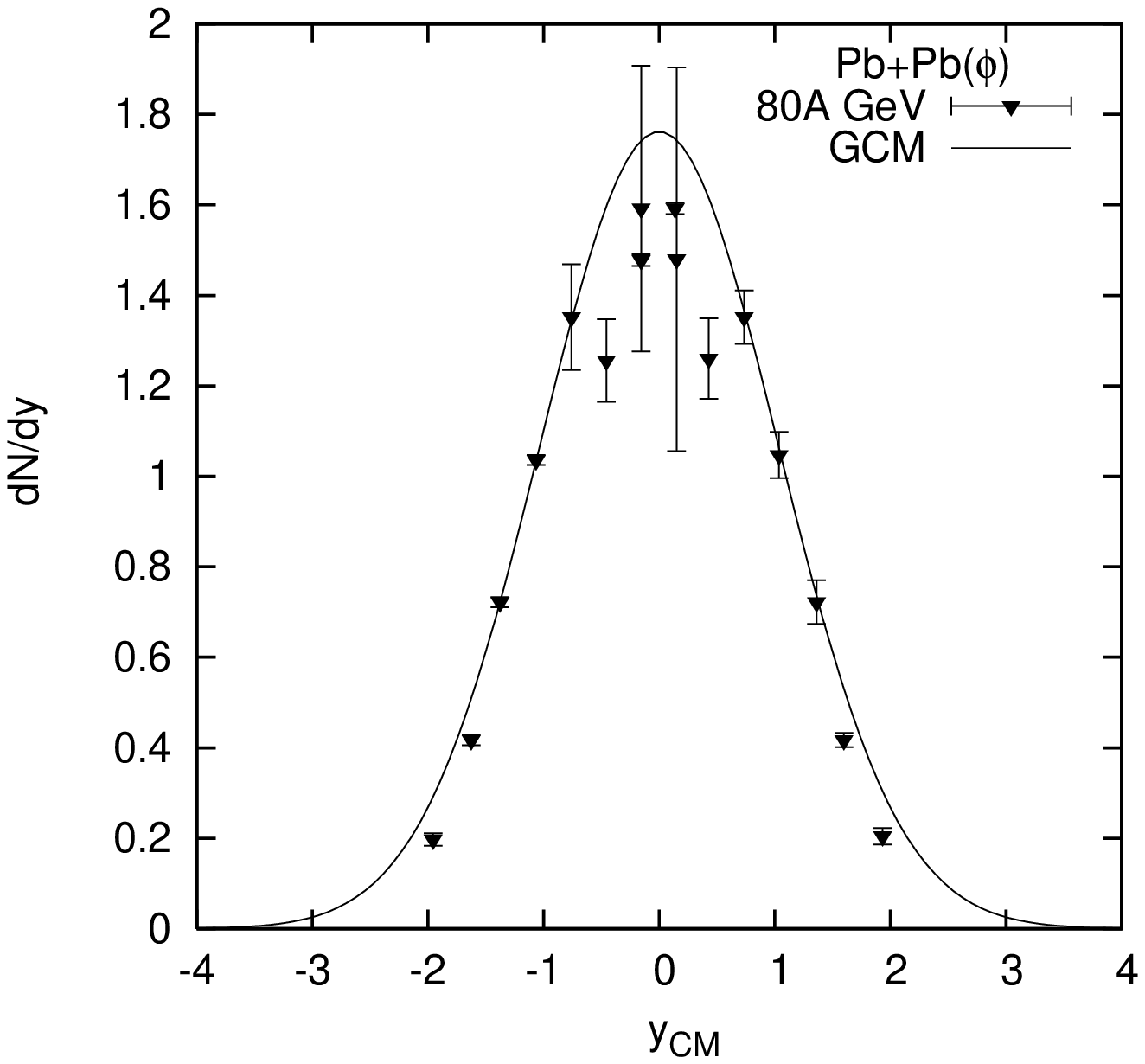}
 \end{minipage}}%
\caption{Rapidity spectra for $\phi$ in central Pb+Pb collisions at $E_{beam}$ = 20, 30, 40, 80
AGeV for $\beta$=0.
The different experimental points are taken from {\cite{Xue1}} and the parameter
values are taken from Table 2-Table 5. The solid curve provide the GCM-based results.}
\end{figure}

\begin{figure}
\subfigure[]{
\begin{minipage}{.5\textwidth}
\centering
 \includegraphics[width=2.5in]{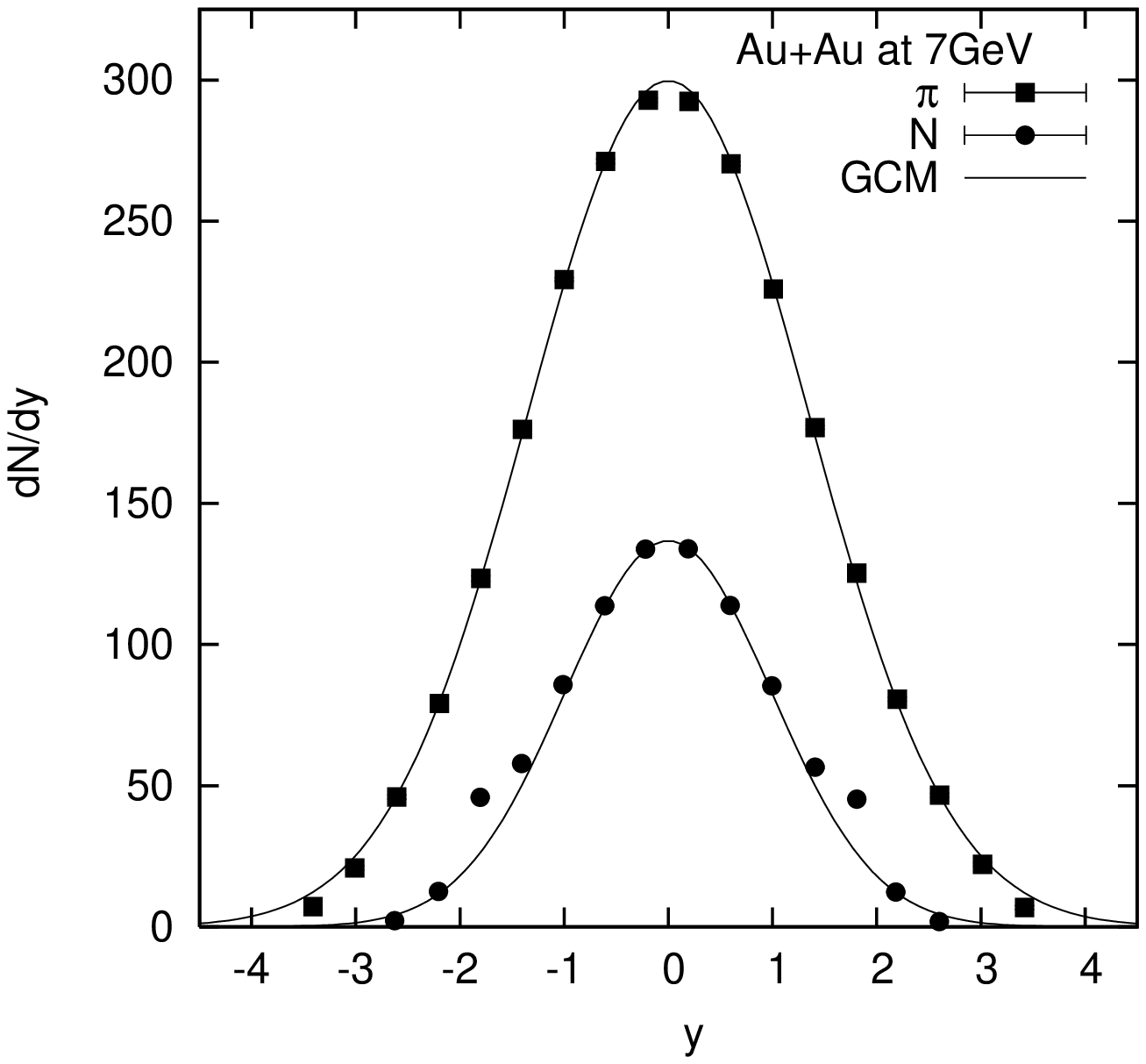}
\end{minipage}}%
\subfigure[]{
\begin{minipage}{0.5\textwidth}
  \centering
\includegraphics[width=2.5in]{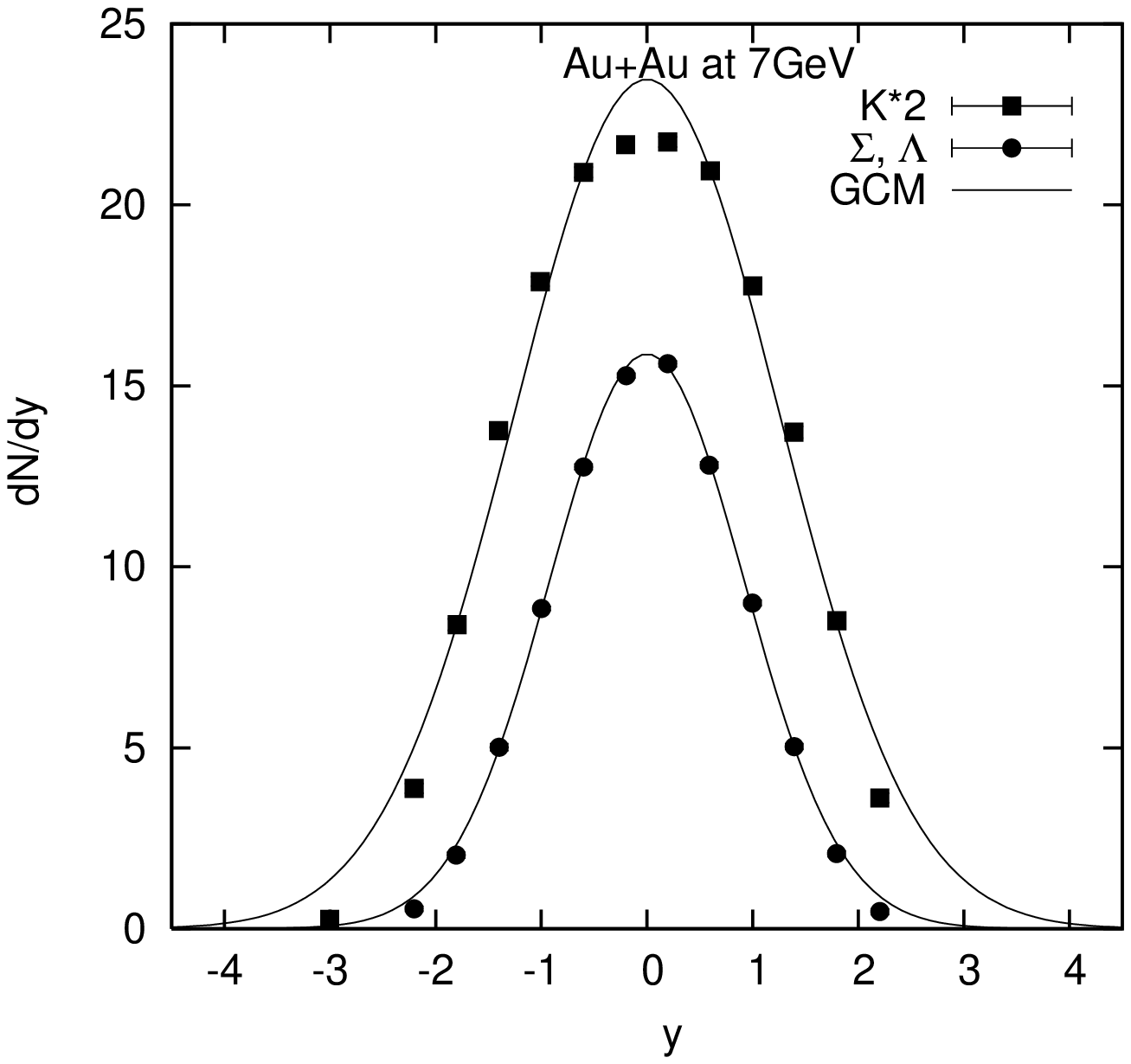}
\end{minipage}}%
\vspace{.01in} \subfigure[]{
\begin{minipage}{1\textwidth}
\centering
 \includegraphics[width=2.5in]{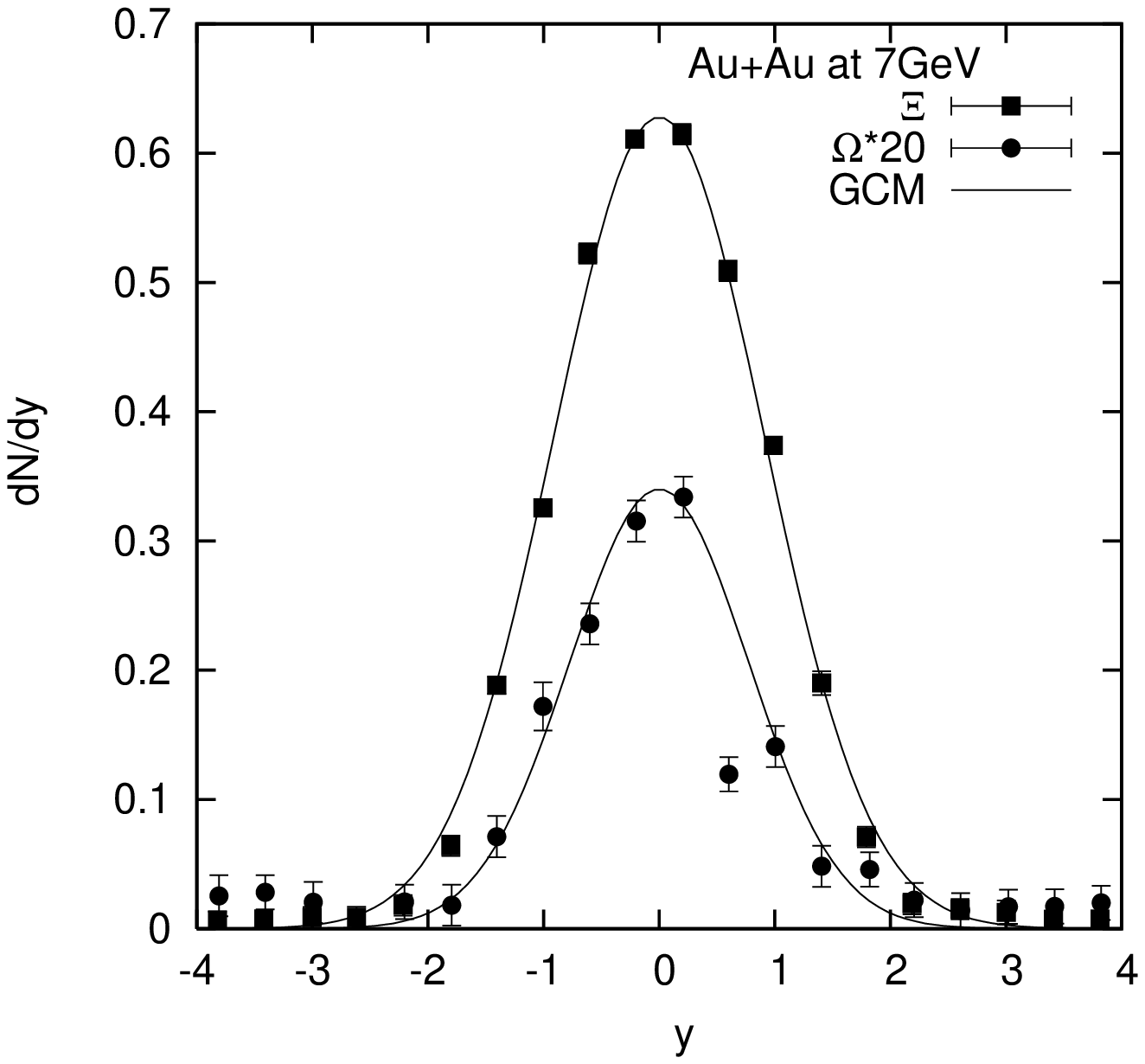}
\end{minipage}}%
\caption{Rapidity distributions of identified hadrons in central Au+Au collisions at $\sqrt{s_{NN}}$=7 GeV for $\beta$=0.
The different experimental points are taken from {\cite{Chen1}} and the parameter
values are taken from Table 6. The solid curve provide the GCM-based results.}
\end{figure}

\begin{figure}
\subfigure[]{
\begin{minipage}{.5\textwidth}
\centering
\includegraphics[width=2.5in]{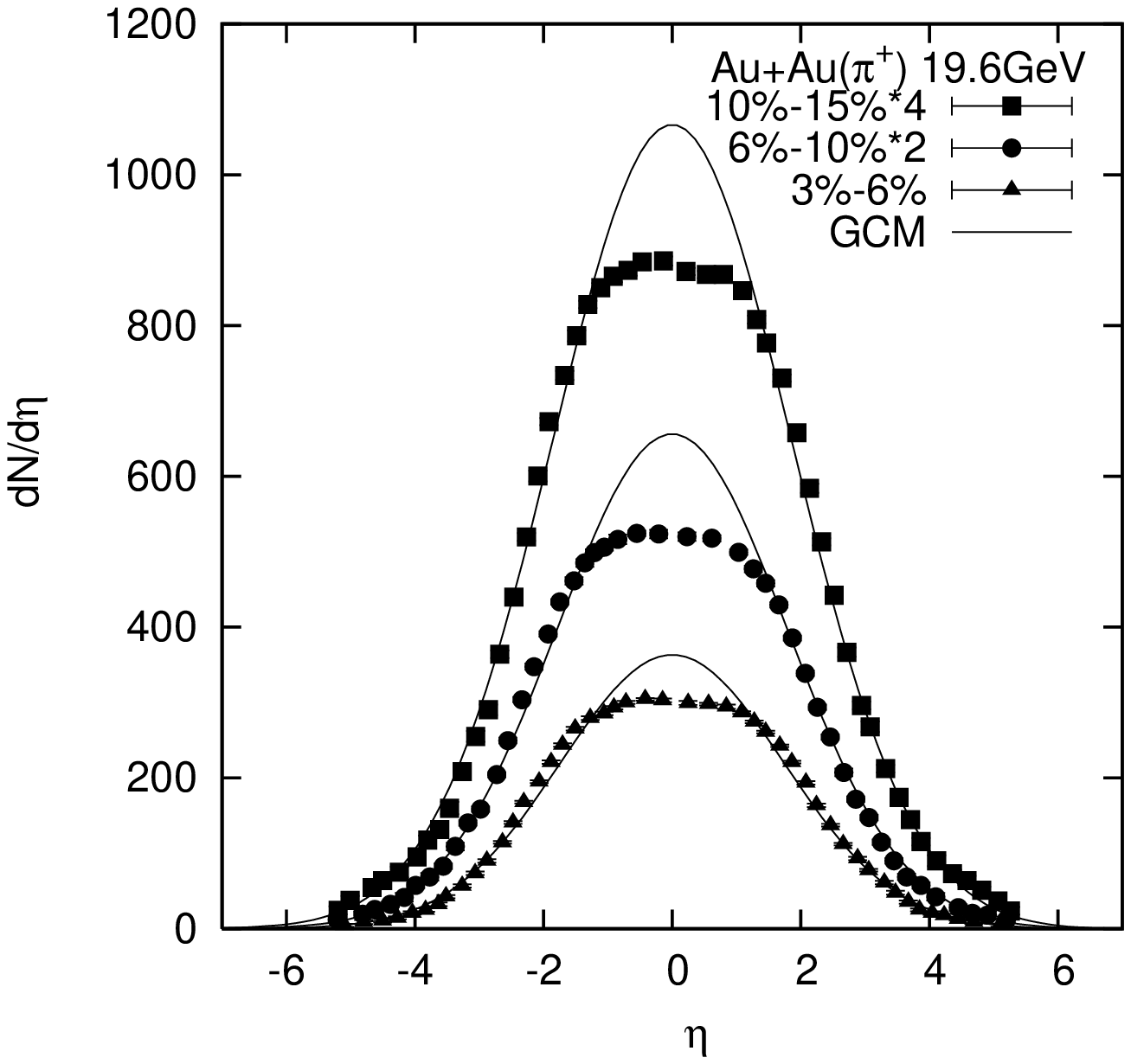}
\setcaptionwidth{2.6in}
\end{minipage}}%
\subfigure[]{
\begin{minipage}{0.5\textwidth}
\centering
 \includegraphics[width=2.5in]{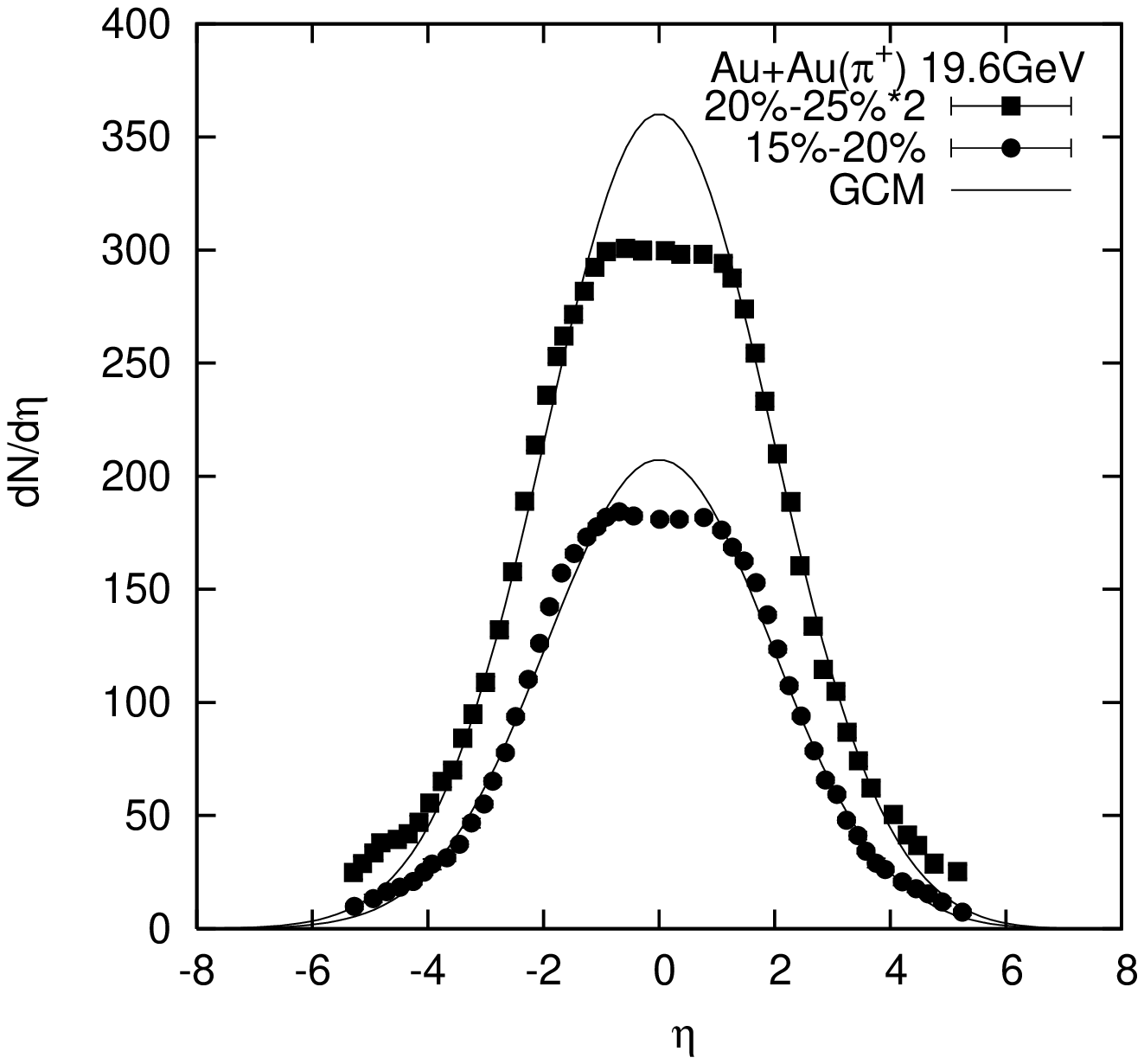}
 \end{minipage}}%
\vspace{0.01in} \subfigure[]{
\begin{minipage}{0.5\textwidth}
\centering
\includegraphics[width=2.5in]{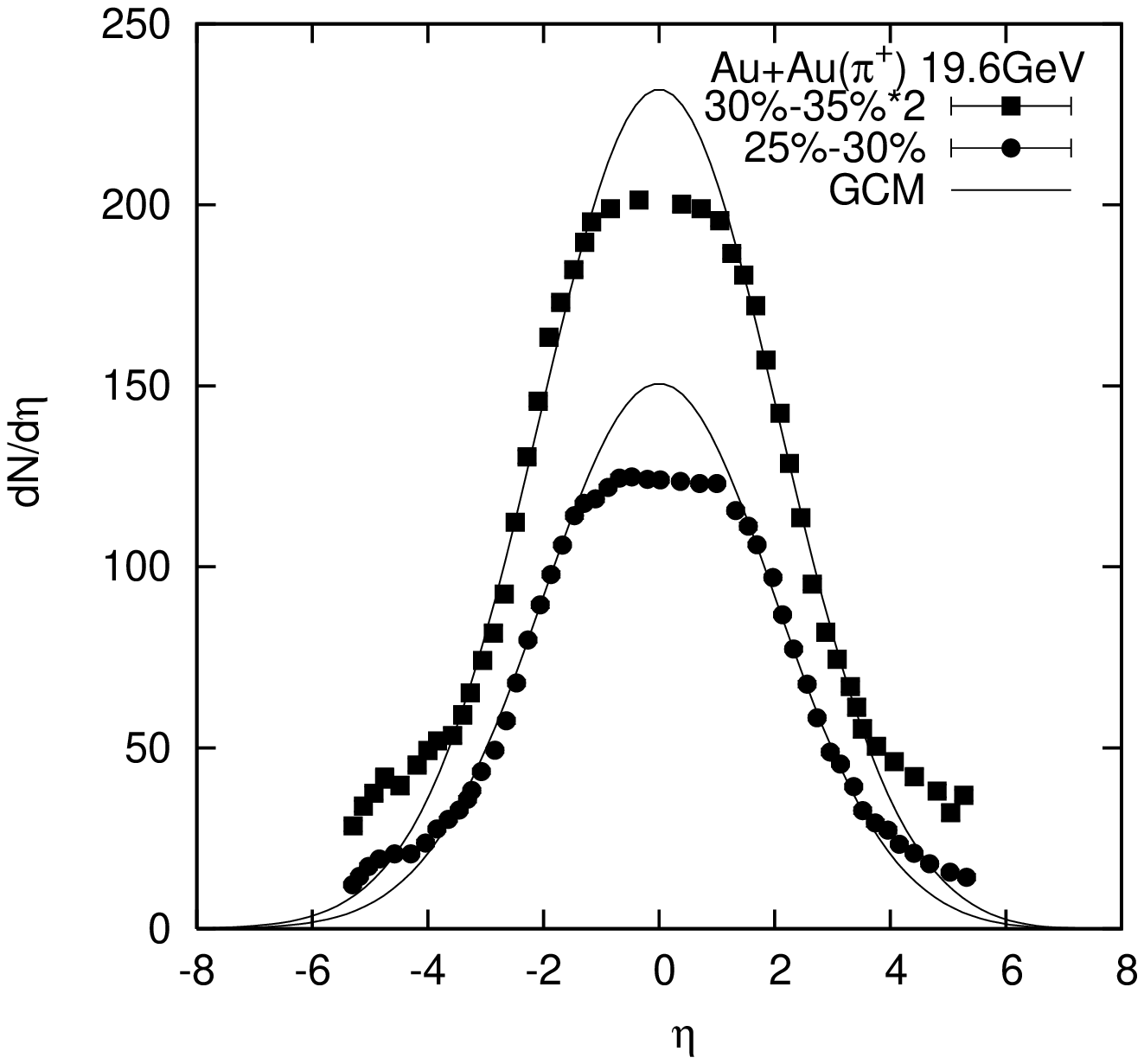}
\end{minipage}}%
\subfigure[]{
\begin{minipage}{.5\textwidth}
\centering
 \includegraphics[width=2.5in]{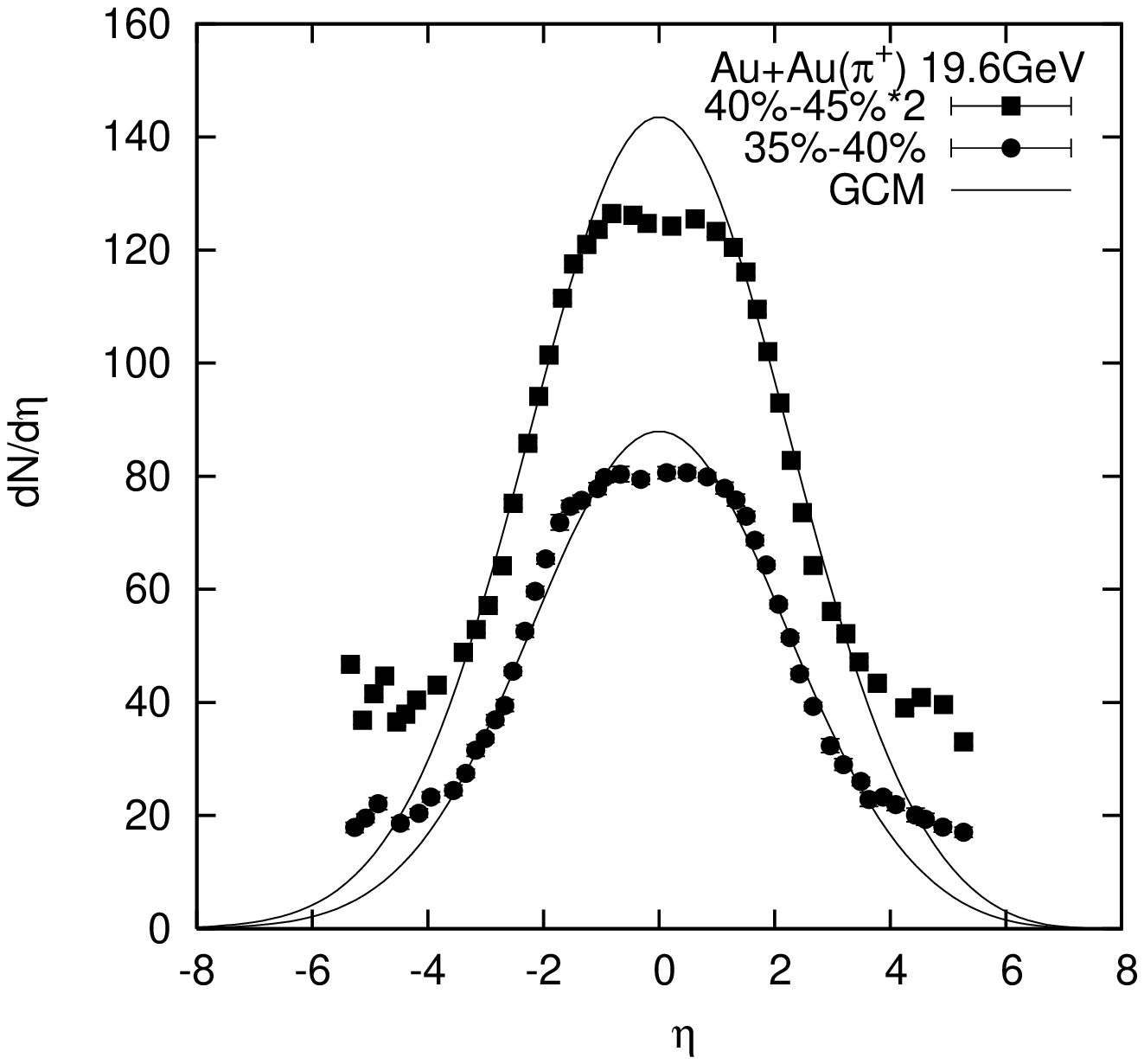}
 \end{minipage}}%
\caption{Pseudo-rapidity spectra for $\pi^+$ for nine centrality bins representing 45\% of the total cross-section
for Au+Au collisions at $\sqrt{s_{NN}}$=19.6 GeV for $\beta$=0.
The different experimental points are taken from {\cite{Alver1}} and the parameter
values are taken from Table 7. The solid curve provide the GCM-based results.}
\end{figure}

\begin{figure}
\subfigure[]{
\begin{minipage}{.5\textwidth}
\centering
 \includegraphics[width=2.5in]{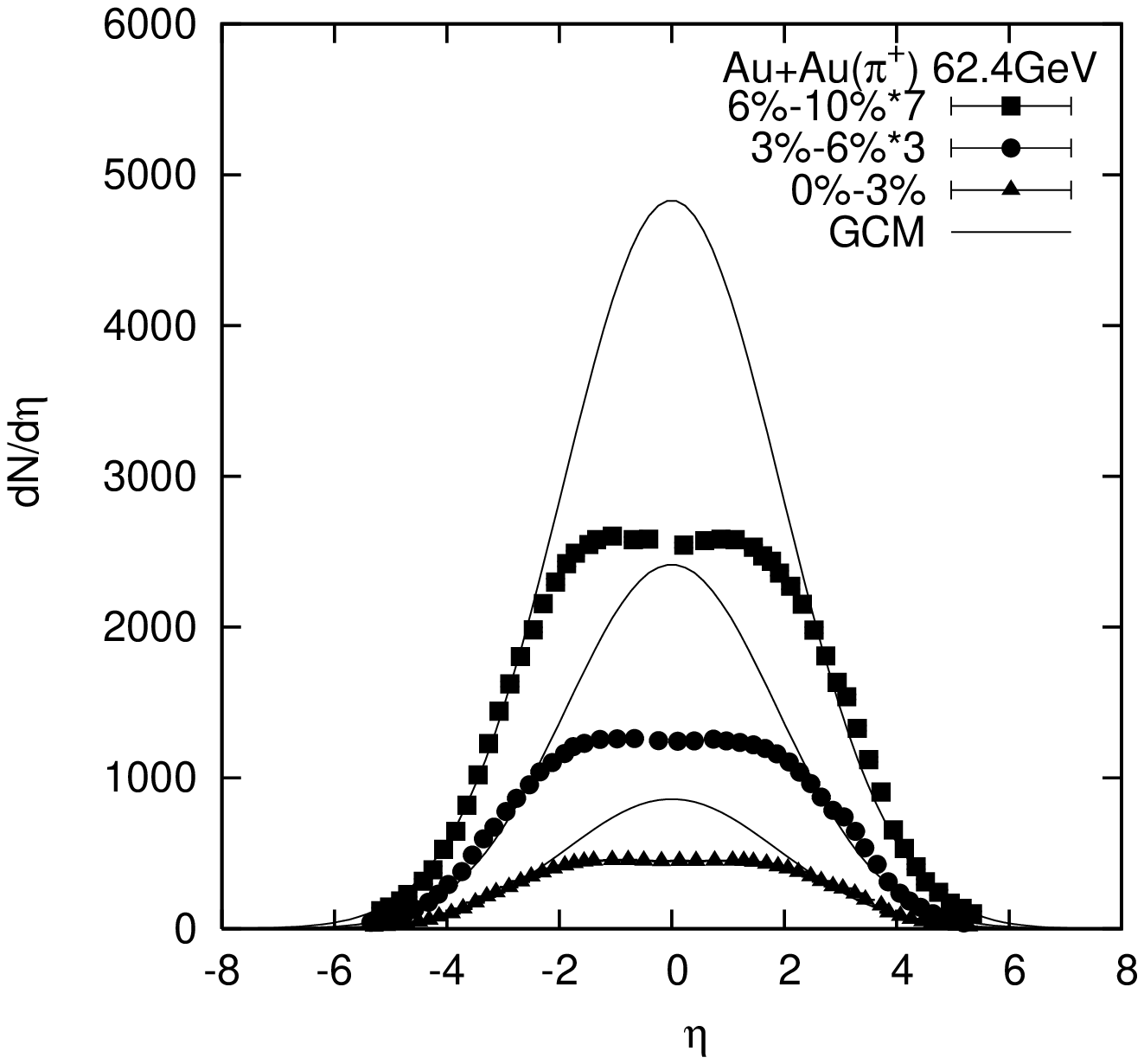}
\end{minipage}}%
\subfigure[]{
\begin{minipage}{0.5\textwidth}
  \centering
\includegraphics[width=2.5in]{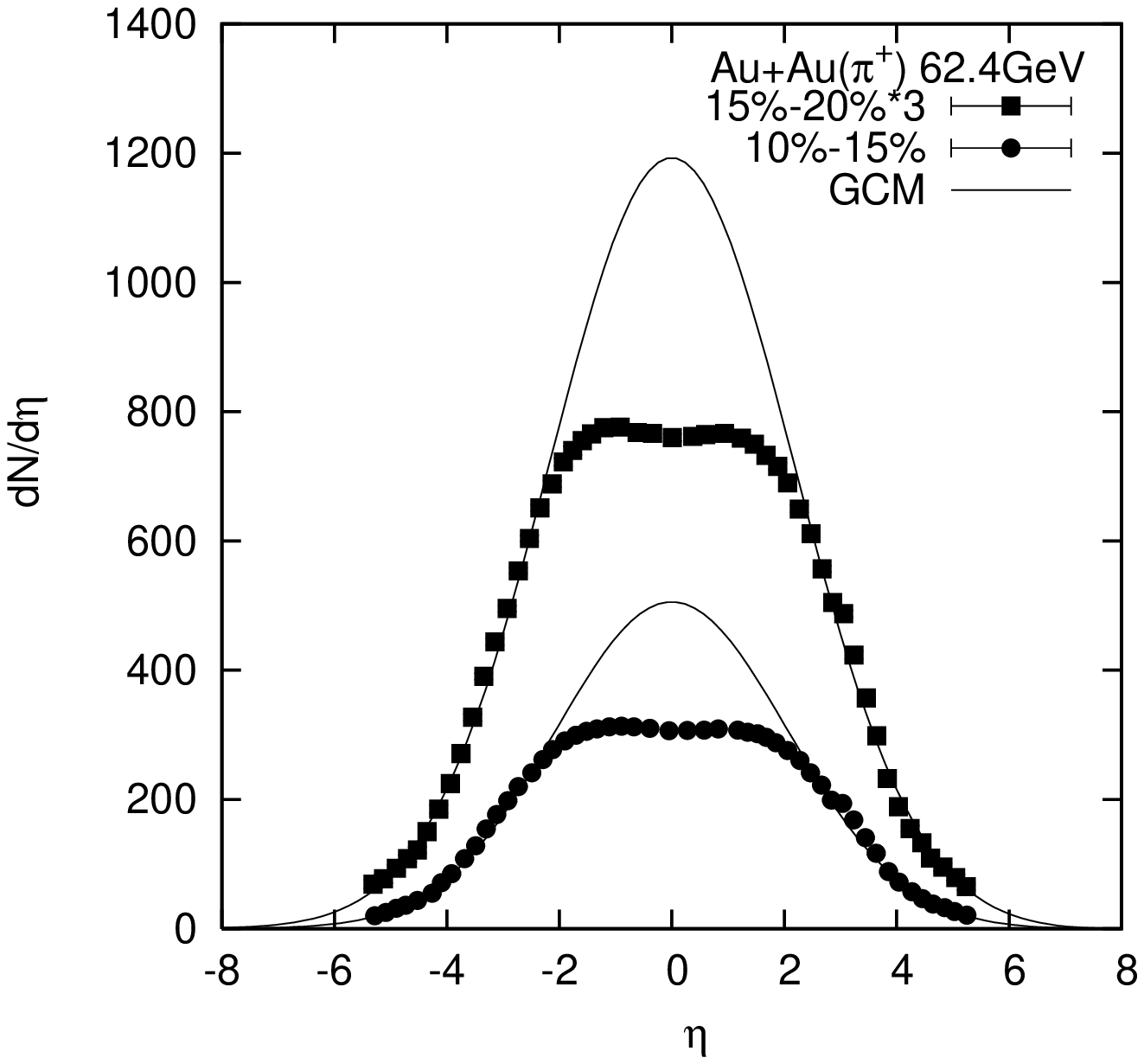}
\end{minipage}}%
\vspace{.01in} \subfigure[]{
\begin{minipage}{1\textwidth}
\centering
 \includegraphics[width=2.5in]{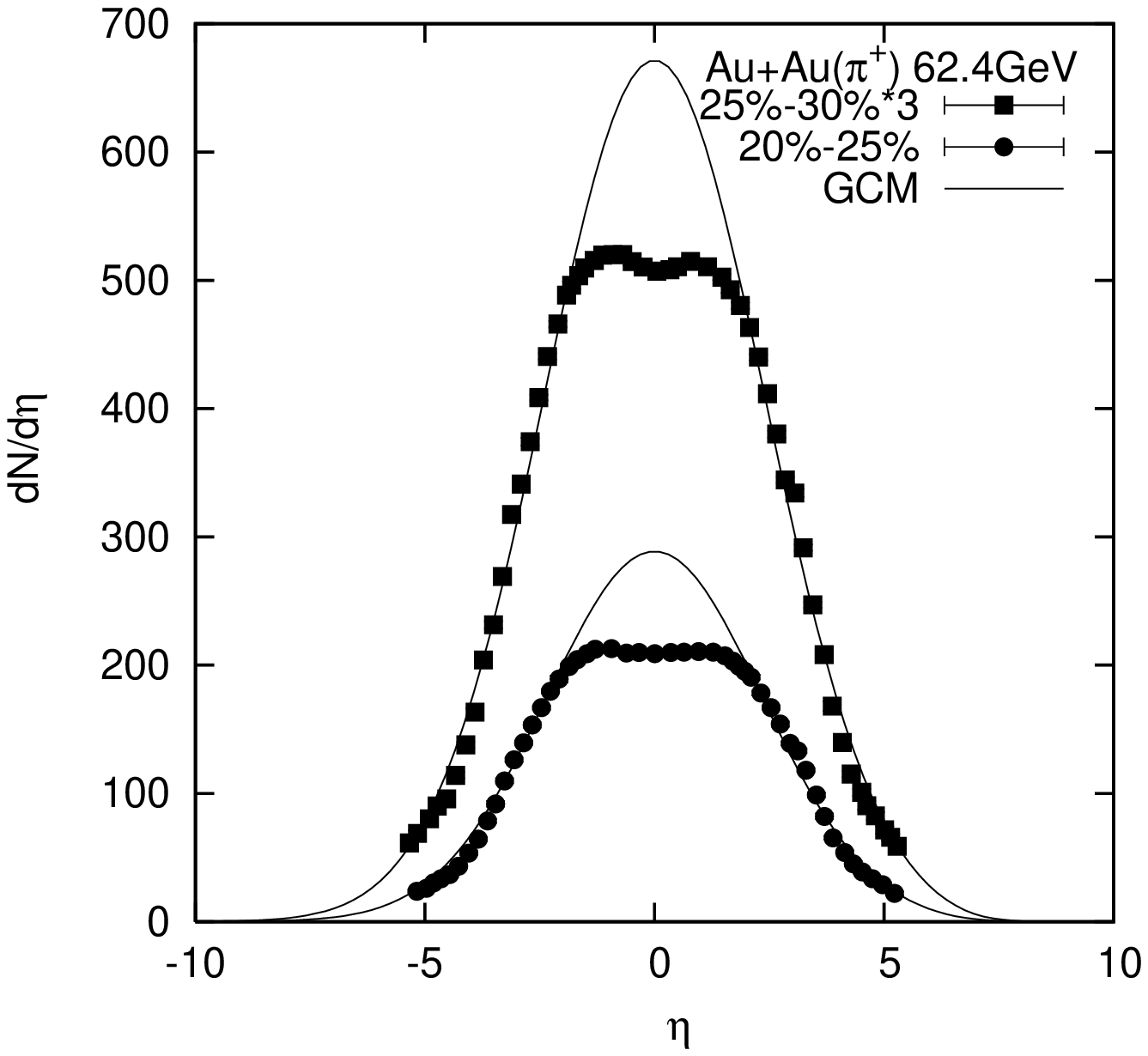}
\end{minipage}}%
\caption{Plot of $\frac{dN}{d\eta}$ vs. $\eta$ for $\pi^+$ for seven centrality bins representing 45\% of
the total cross-section for Au+Au collisions at $\sqrt{s_{NN}}$=62.4 GeV for $\beta$=0.
The different experimental points are taken from {\cite{Alver1}} and the parameter
values are taken from Table 8. The solid curve provide the GCM-based results.}
\end{figure}

\begin{figure}
\subfigure[]{
\begin{minipage}{.5\textwidth}
\centering
\includegraphics[width=2.5in]{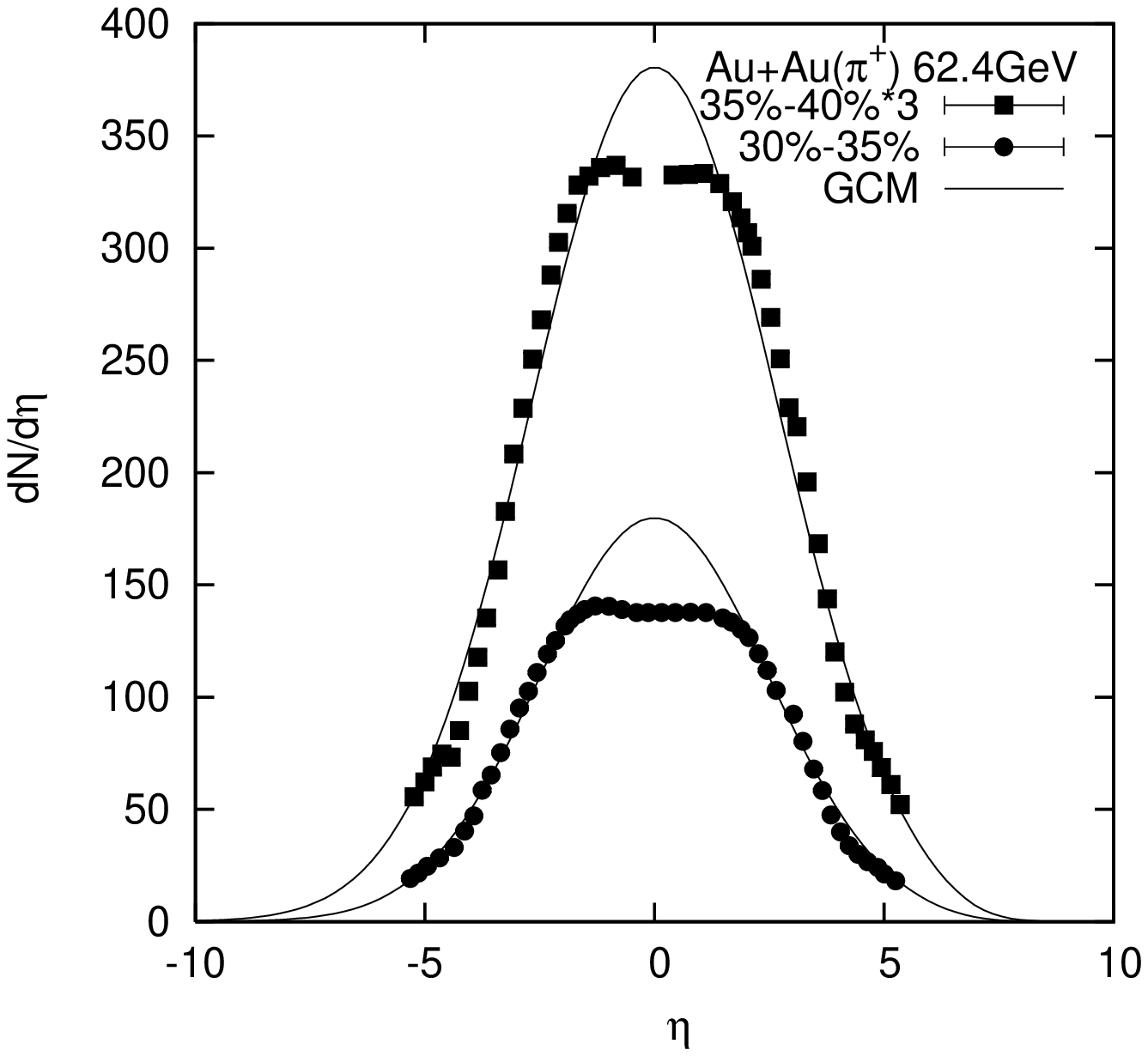}
\end{minipage}}%
\subfigure[]{
\begin{minipage}{.5\textwidth}
\centering
 \includegraphics[width=2.5in]{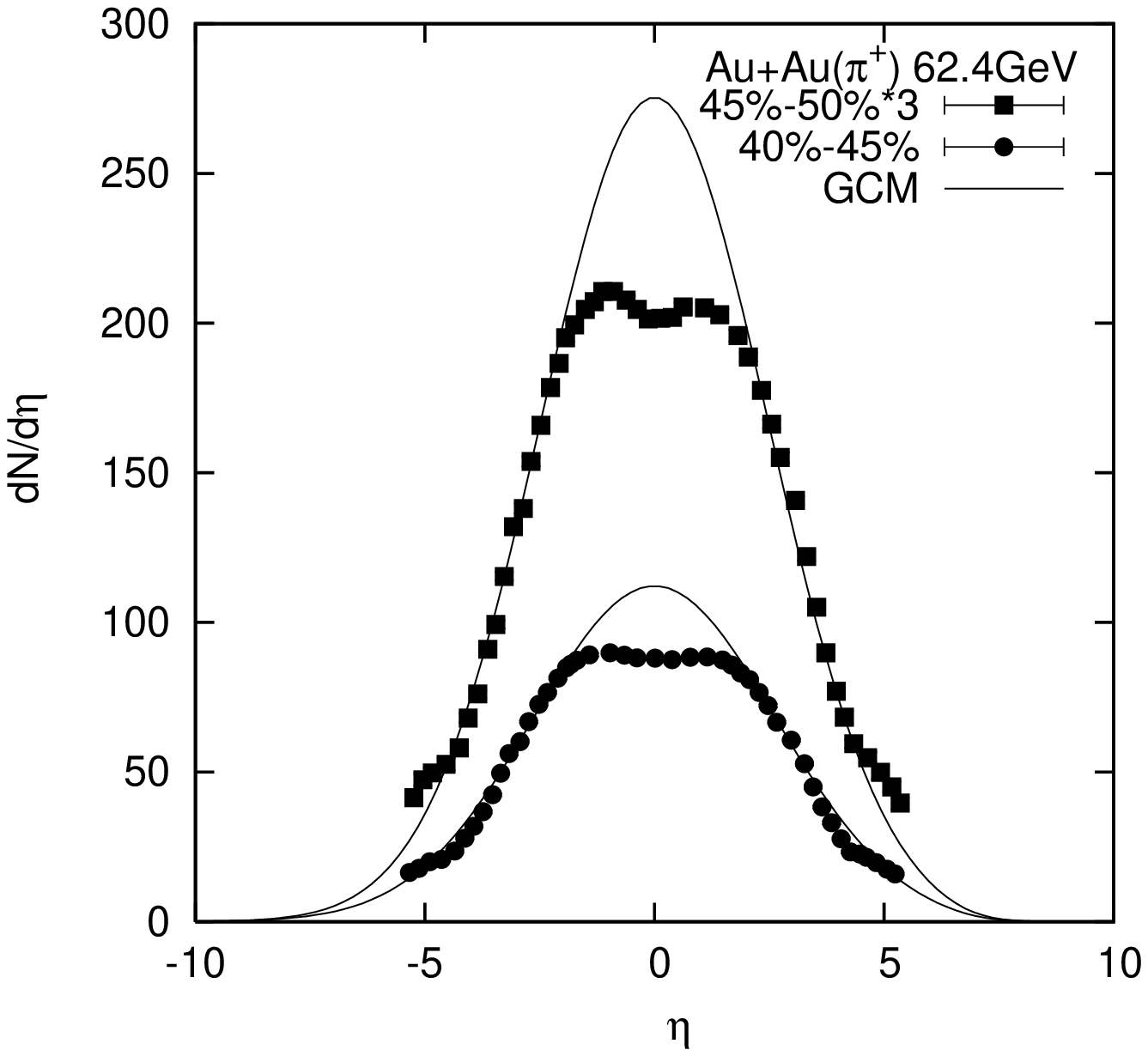}
 \end{minipage}}%
\caption{Plot of $\frac{dN}{d\eta}$ vs. $\eta$ for $\pi^+$ for four centrality bins representing 45\% of
the total cross-section for Au+Au collisions at $\sqrt{s_{NN}}$=62.4 GeV for $\beta$=0.
The different experimental points are taken from {\cite{Alver1}} and the parameter
values are taken from Table 8. The solid curve provide the GCM-based results.}

\subfigure[]{
\begin{minipage}{.5\textwidth}
\centering
\includegraphics[width=2.5in]{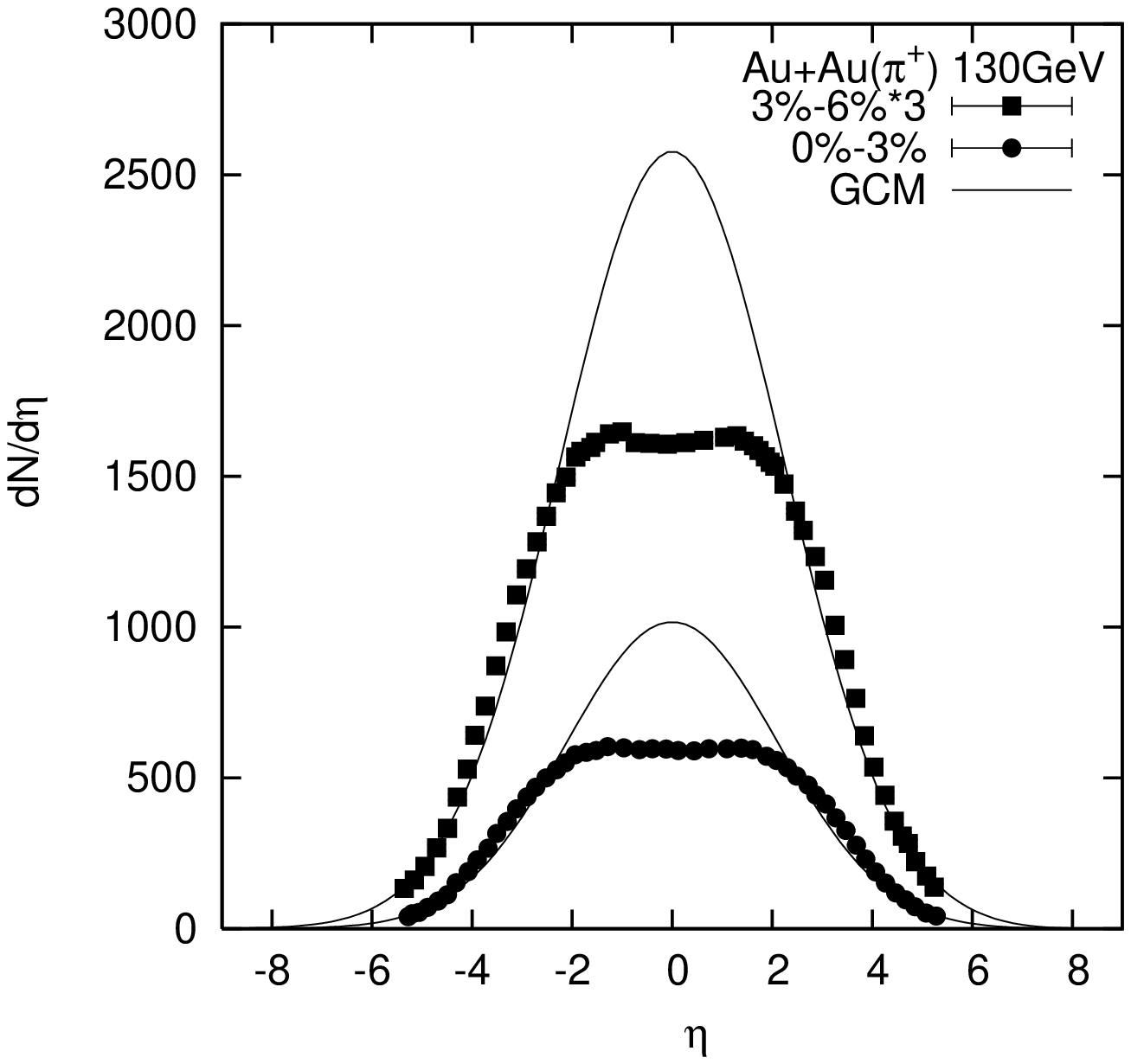}
\end{minipage}}%
\subfigure[]{
\begin{minipage}{.5\textwidth}
\centering
 \includegraphics[width=2.5in]{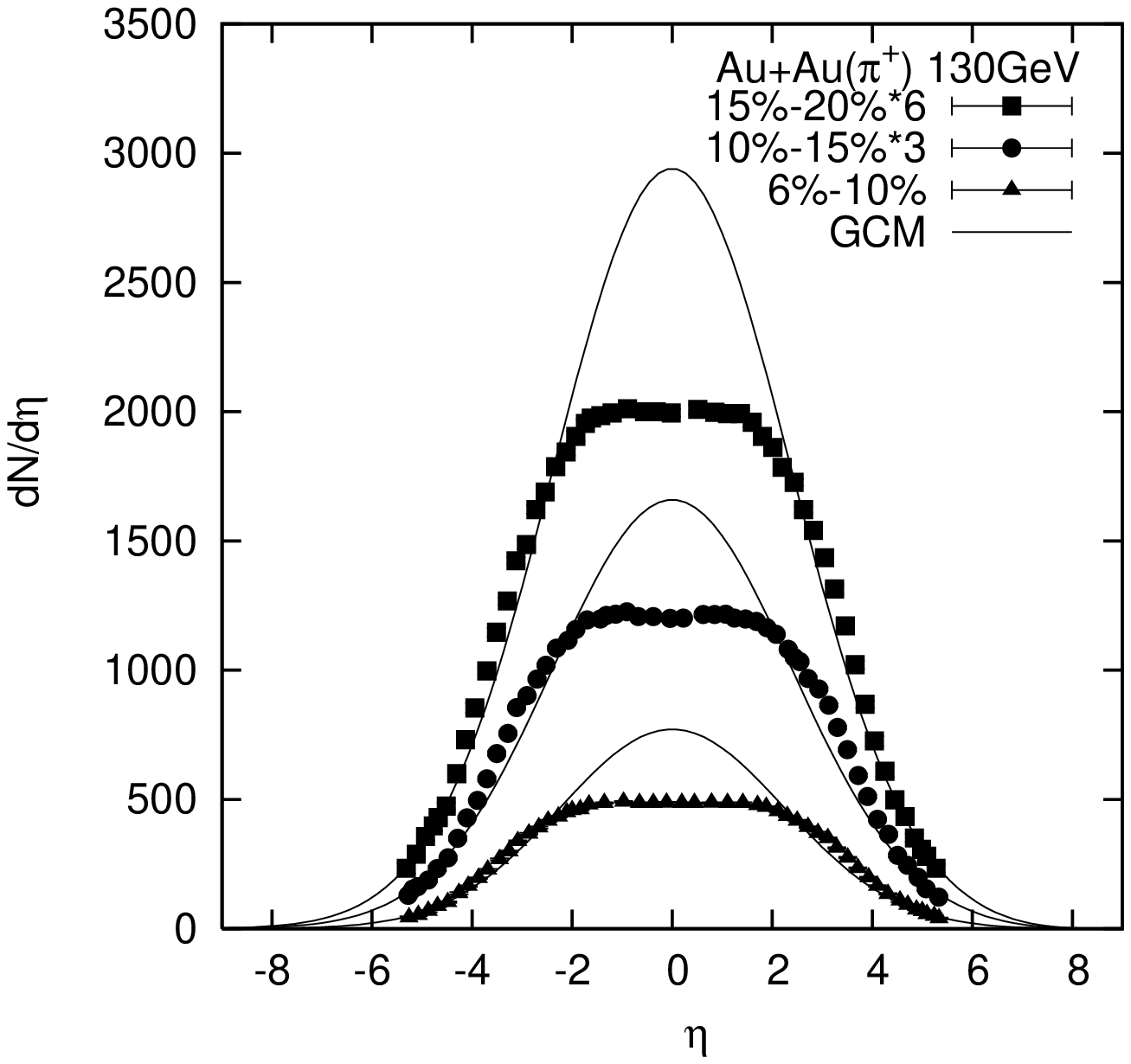}
 \end{minipage}}%
\caption{Pseudo-rapidity spectra for $\pi^+$ for five centrality bins representing 45\% of the total cross-section
for Au+Au collisions at $\sqrt{s_{NN}}$=130 GeV for $\beta$=0.
The different experimental points are taken from {\cite{Alver1}} and the parameter
values are taken from Table 9. The solid curve provide the GCM-based results.}
\end{figure}

\begin{figure}
\subfigure[]{
\begin{minipage}{.5\textwidth}
\centering
 \includegraphics[width=2.5in]{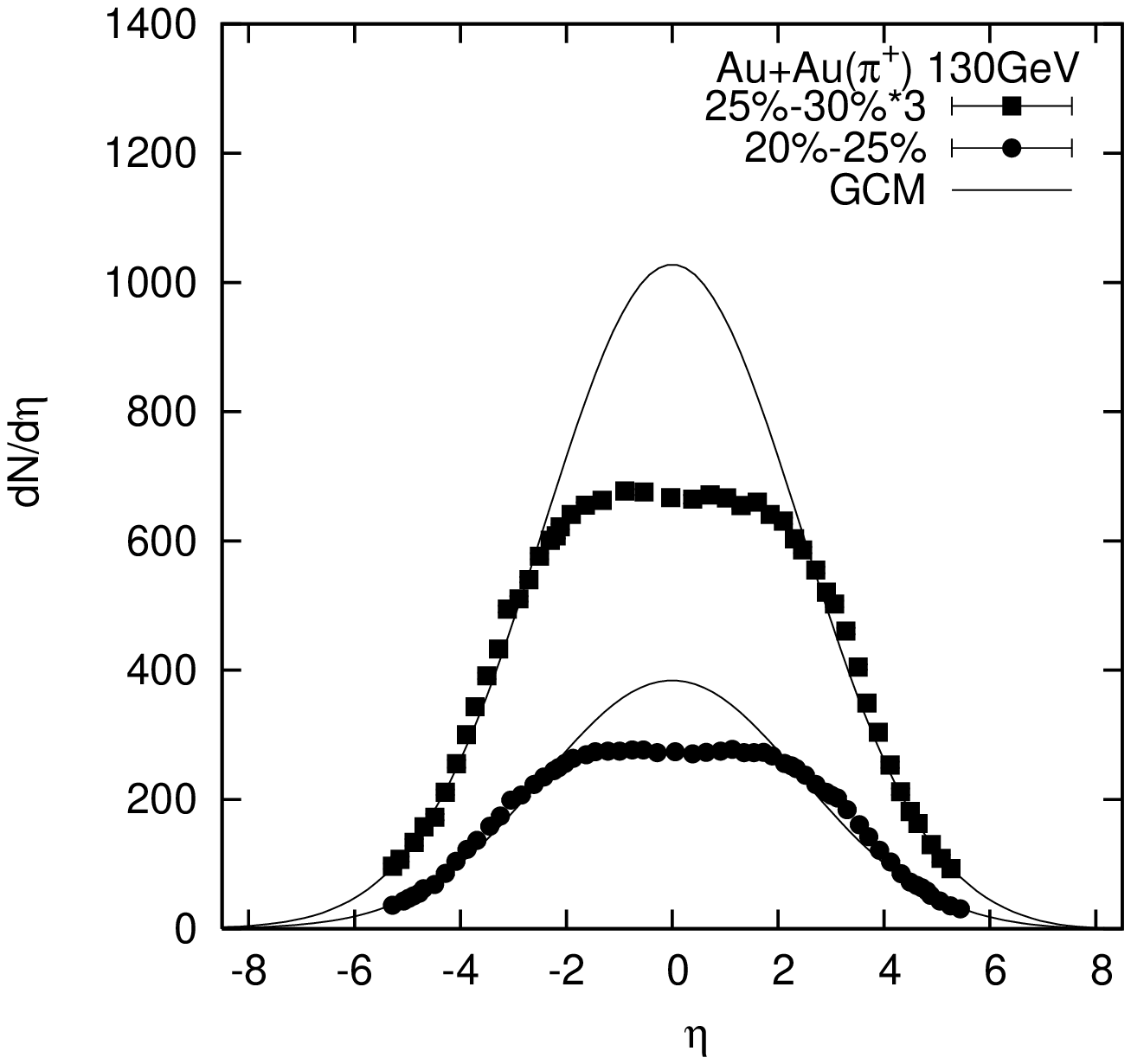}
\end{minipage}}%
\subfigure[]{
\begin{minipage}{0.5\textwidth}
  \centering
\includegraphics[width=2.5in]{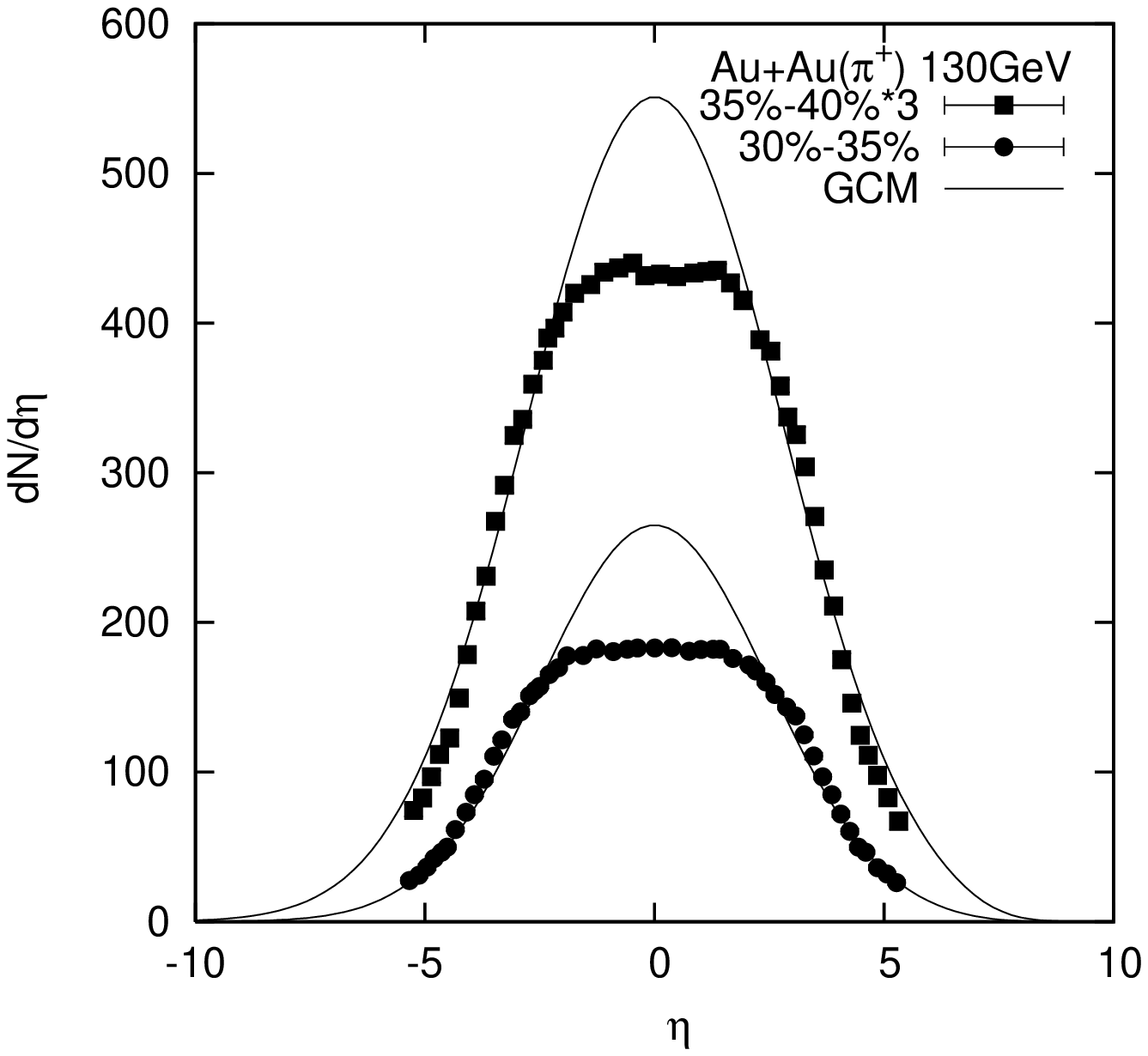}
\end{minipage}}%
\vspace{.01in} \subfigure[]{
\begin{minipage}{1\textwidth}
\centering
 \includegraphics[width=2.5in]{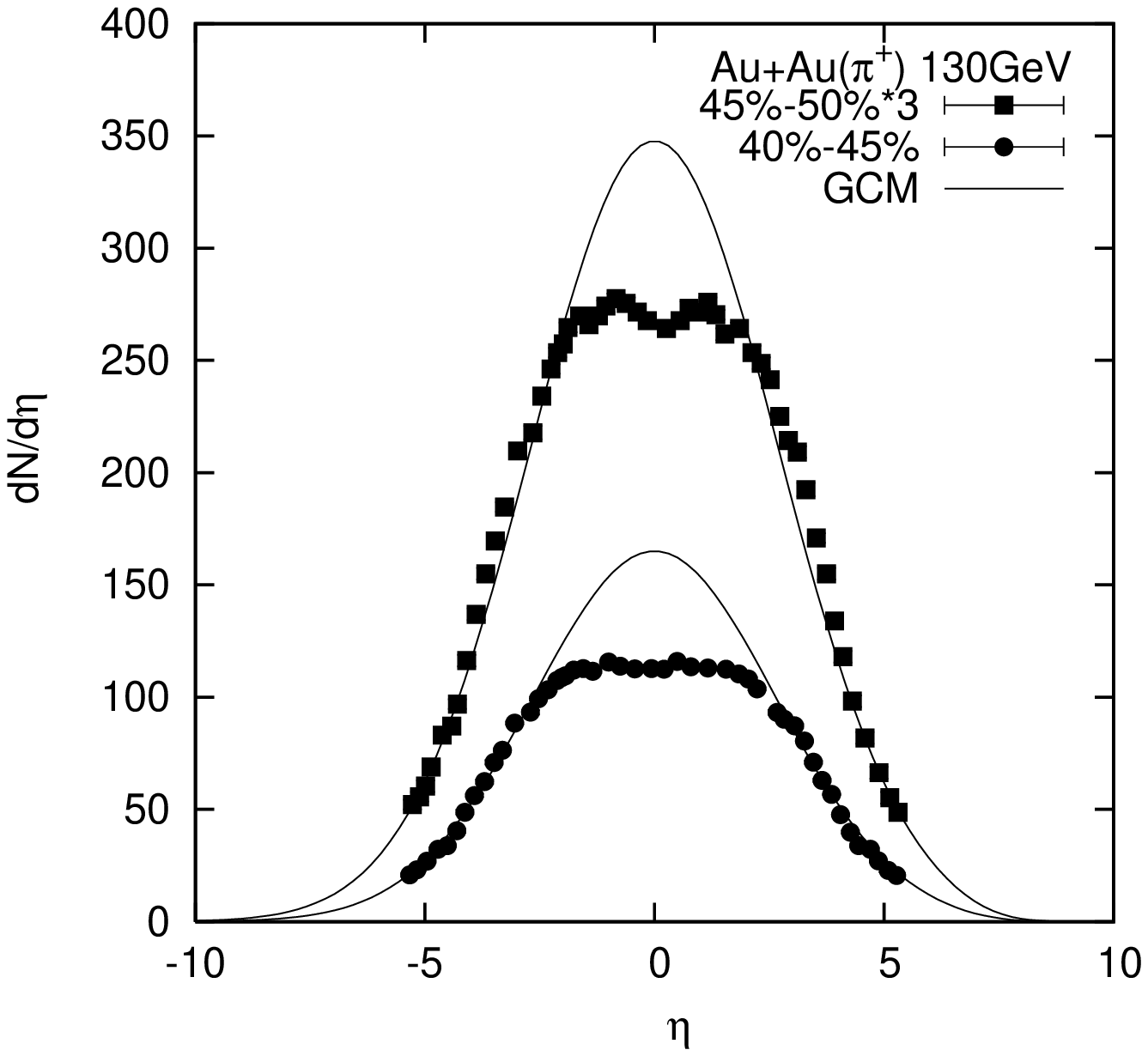}
\end{minipage}}%
\caption{Pseudo-rapidity spectra for $\pi^+$ for six centrality bins representing 45\% of the total cross-section
for Au+Au collisions at $\sqrt{s_{NN}}$=130 GeV for $\beta$=0.
The different experimental points are taken from {\cite{Alver1}} and the parameter
values are taken from Table 9. The solid curve provide the GCM-based results.}
\end{figure}

\begin{figure}
\subfigure[]{
\begin{minipage}{.5\textwidth}
\centering
 \includegraphics[width=2.5in]{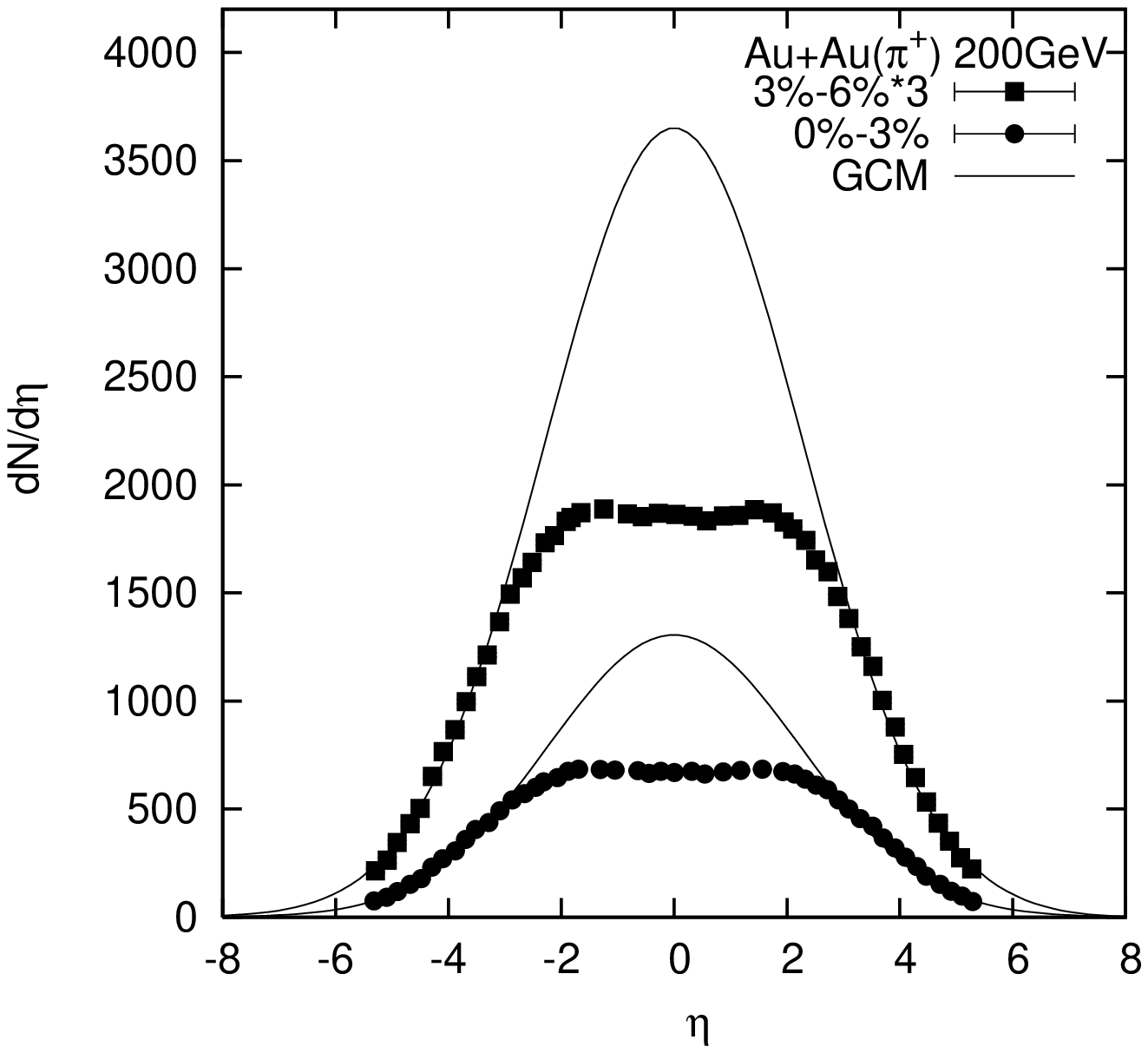}
\end{minipage}}%
\subfigure[]{
\begin{minipage}{0.5\textwidth}
  \centering
\includegraphics[width=2.5in]{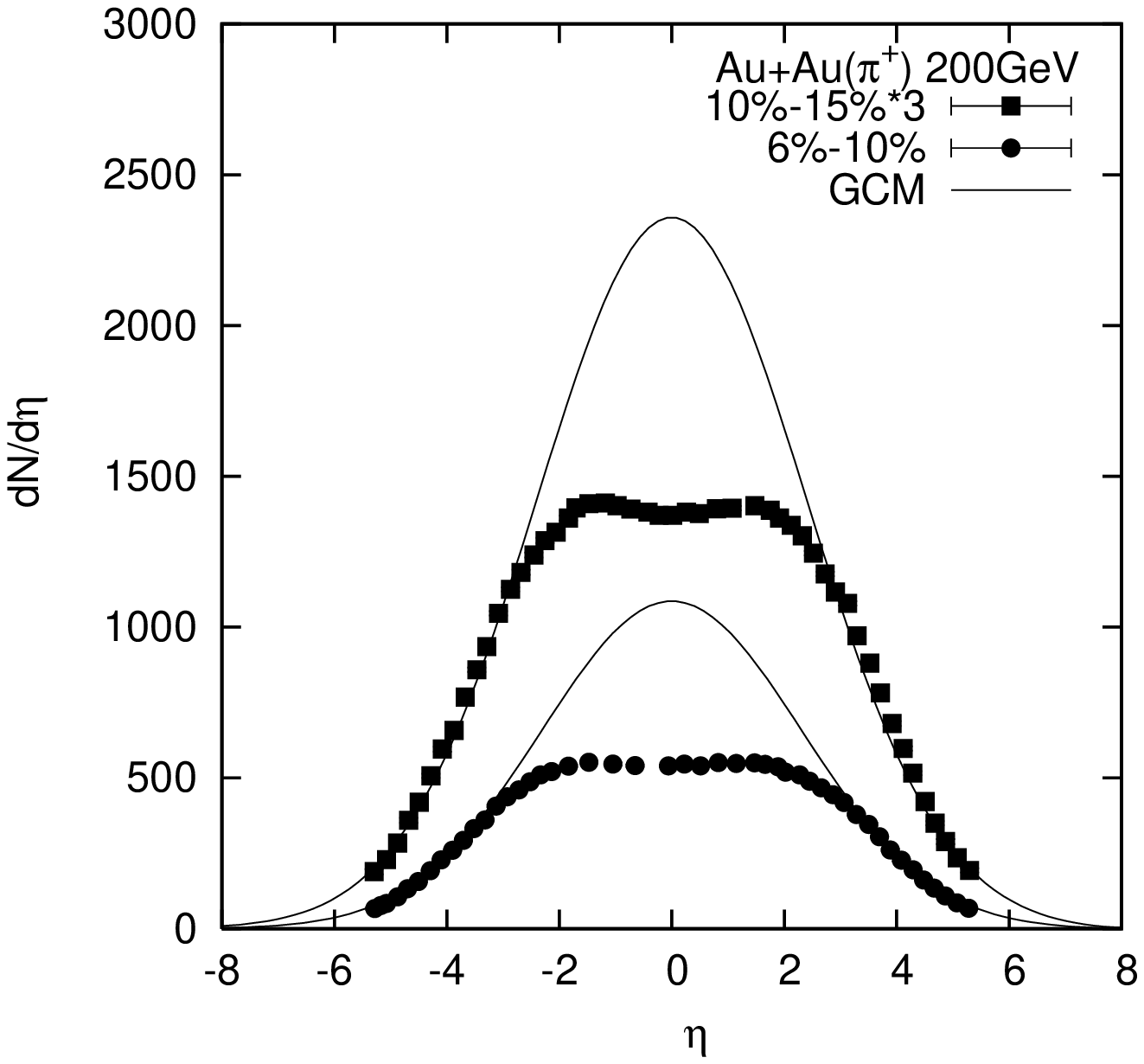}
\end{minipage}}%
\vspace{.01in} \subfigure[]{
\begin{minipage}{1\textwidth}
\centering
 \includegraphics[width=2.5in]{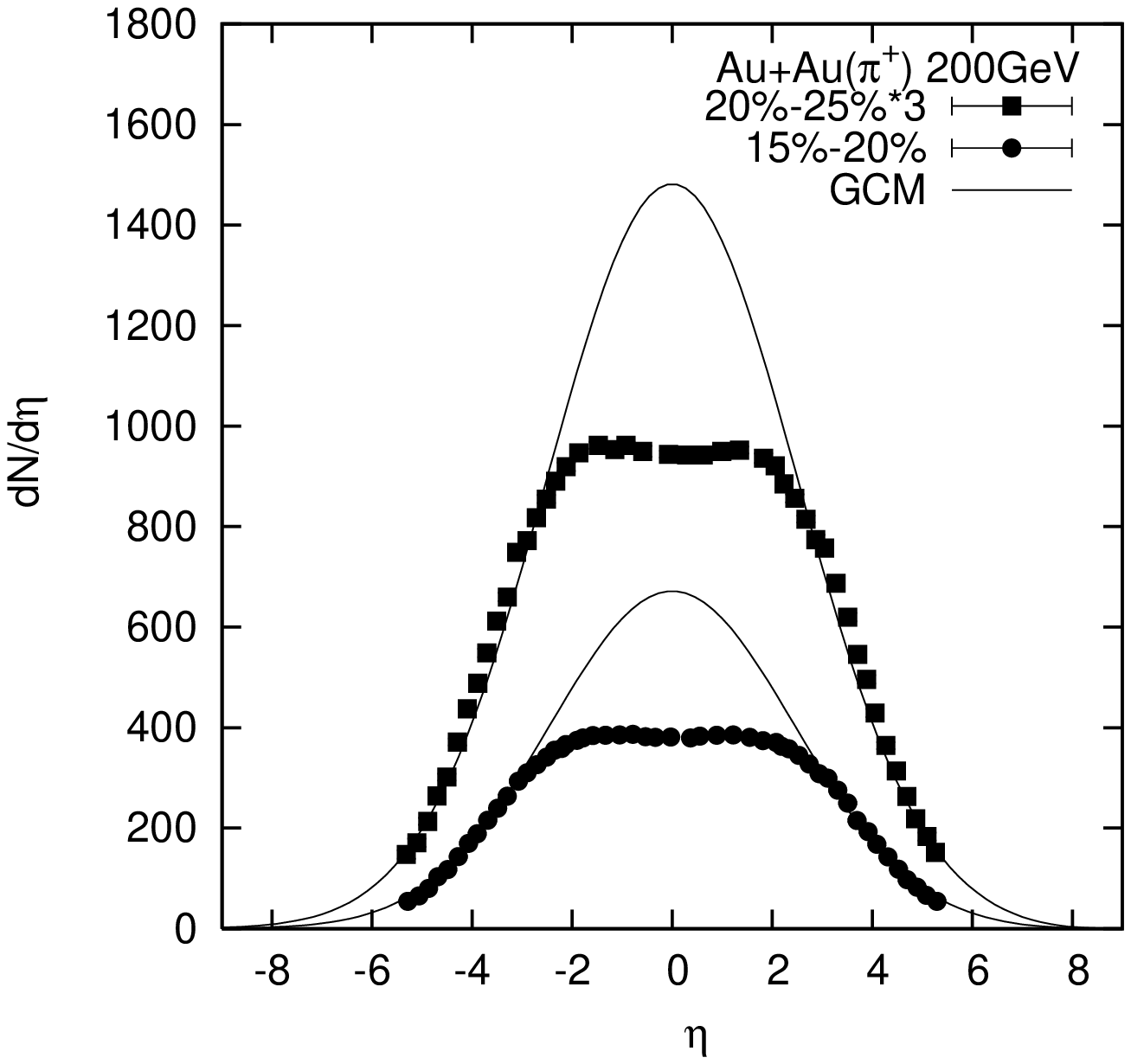}
\end{minipage}}%
\caption{Plot of $\frac{dN}{d\eta}$ vs. $\eta$ for $\pi^+$ for six centrality bins representing 45\%
of the total cross-section for Au+Au collisions at $\sqrt{s_{NN}}$=200 GeV for $\beta$=0.
The different experimental points are taken from {\cite{Alver1}} and the parameter
values are taken from Table 10. The solid curve provide the GCM-based results.}
\end{figure}

\begin{figure}
\subfigure[]{
\begin{minipage}{.5\textwidth}
\centering
\includegraphics[width=2.5in]{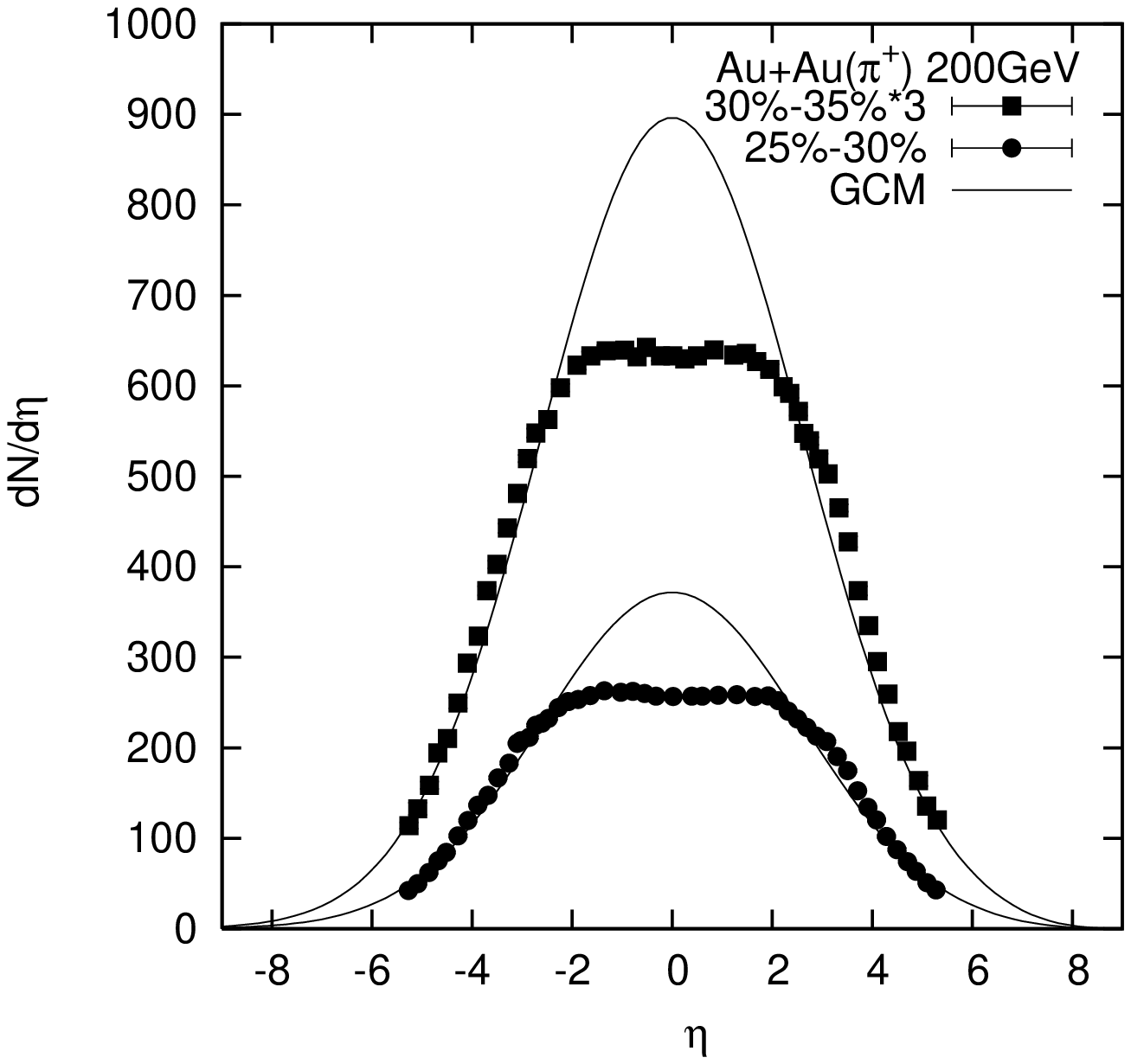}
\end{minipage}}%
\subfigure[]{
\begin{minipage}{.5\textwidth}
\centering
 \includegraphics[width=2.5in]{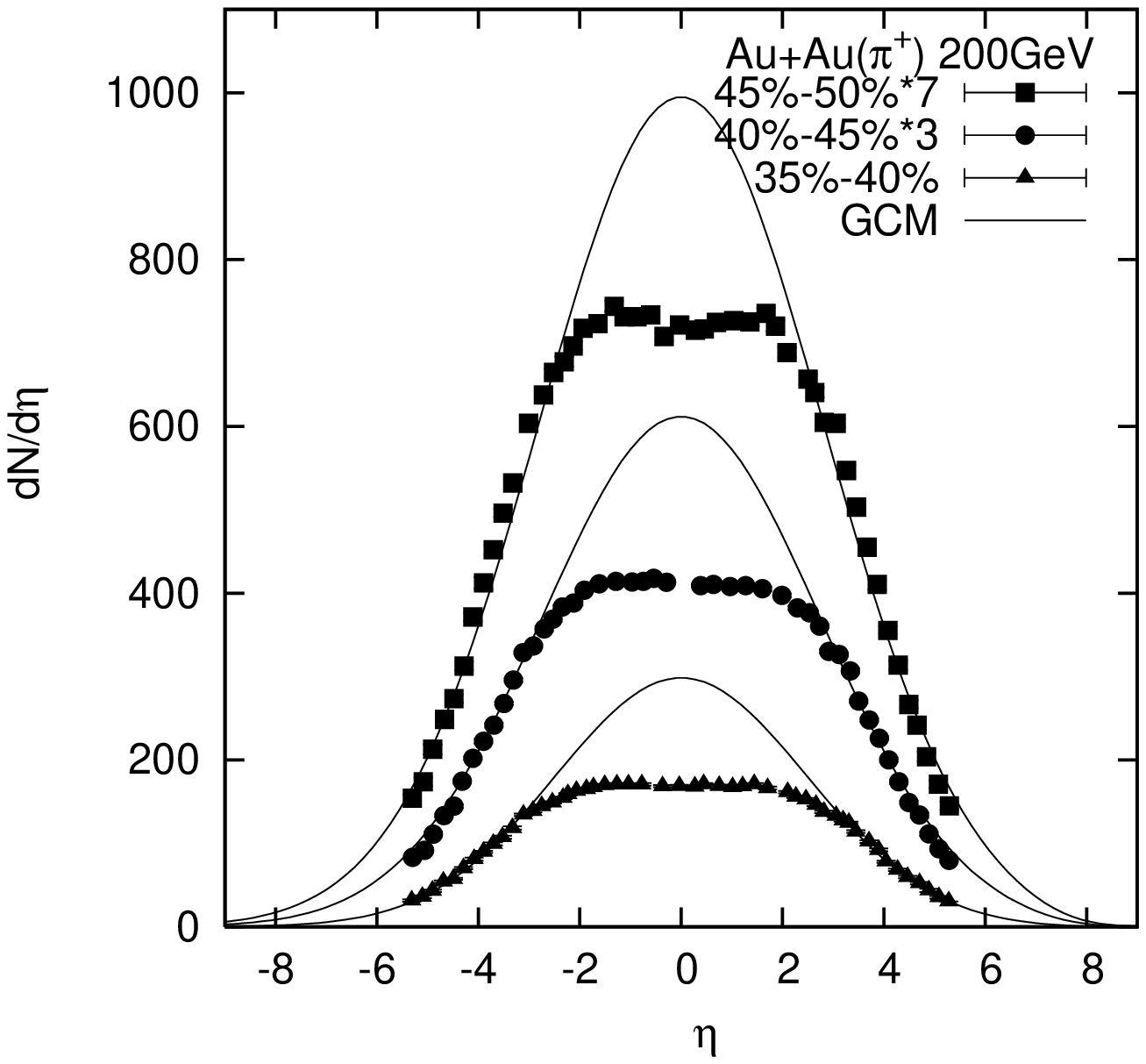}
 \end{minipage}}%
\caption{Plot of $\frac{dN}{d\eta}$ vs. $\eta$ for $\pi^+$ for five centrality bins representing
 45\% of the total cross-section for Au+Au collisions at $\sqrt{s_{NN}}$=200 GeV for $\beta$=0.
The different experimental points are taken from {\cite{Alver1}} and the parameter
values are taken from Table 10. The solid curve provide the GCM-based results.}
\end{figure}

\begin{figure} \centering
\includegraphics[width=2.5in]{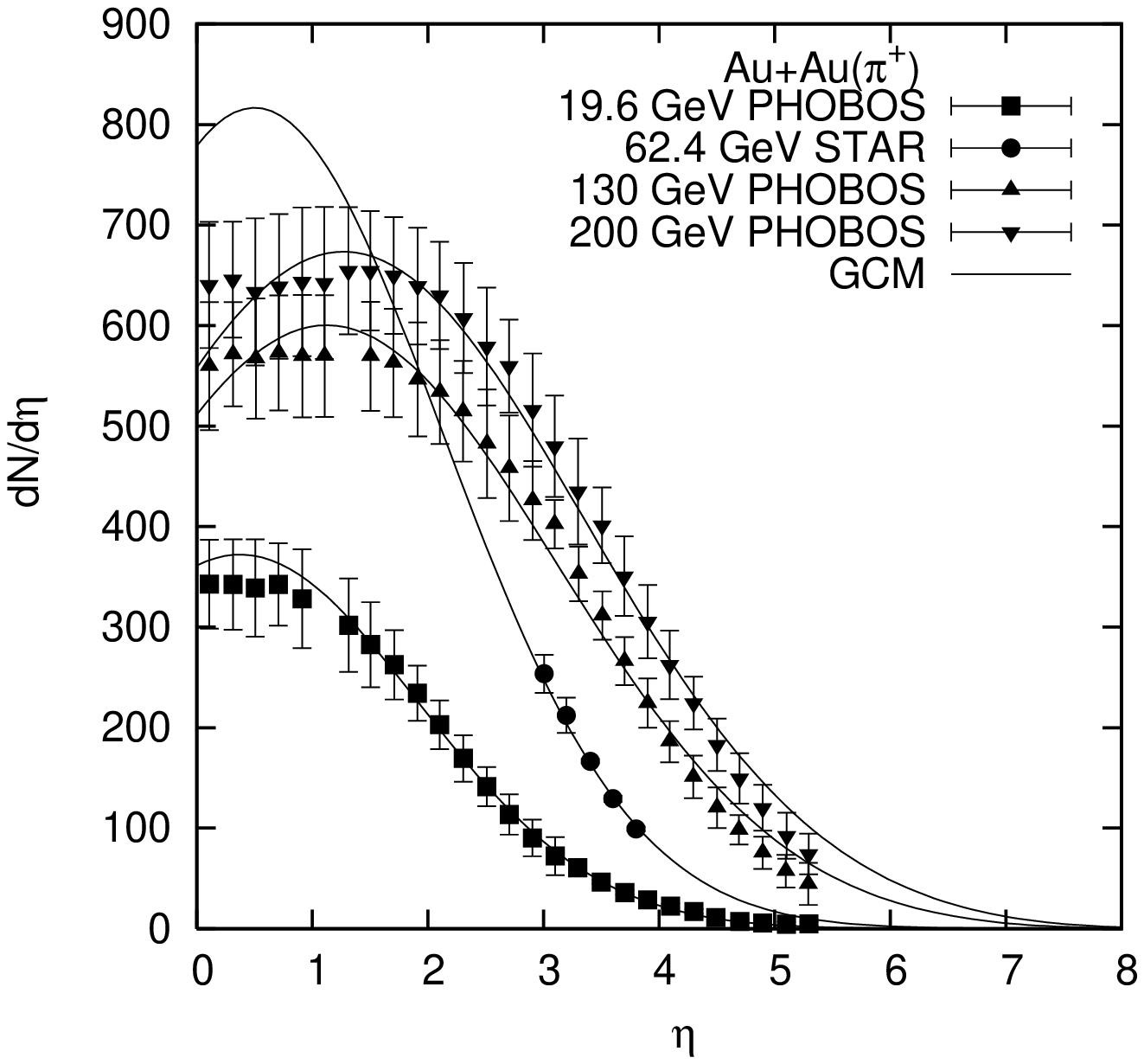}
\caption{Pseudorapidity distributions of charged
particles(for $\pi^+$) for various c.m. energies in Au+Au central collisions.
Pseudorapidity distributions for 0-6\% central Au+Au collisions at
$\sqrt{s_{NN}}$ = 200, 130, and 19.6 GeV are from the PHOBOS experiment and 62.4 GeV is from STAR experiment($\beta$$\neq$0).
The different experimental points are taken from {\cite{Adams1}} and the parameter
values are taken from Table 11 . The solid curve provide the GCM-based results.}

\includegraphics[width=2.5in]{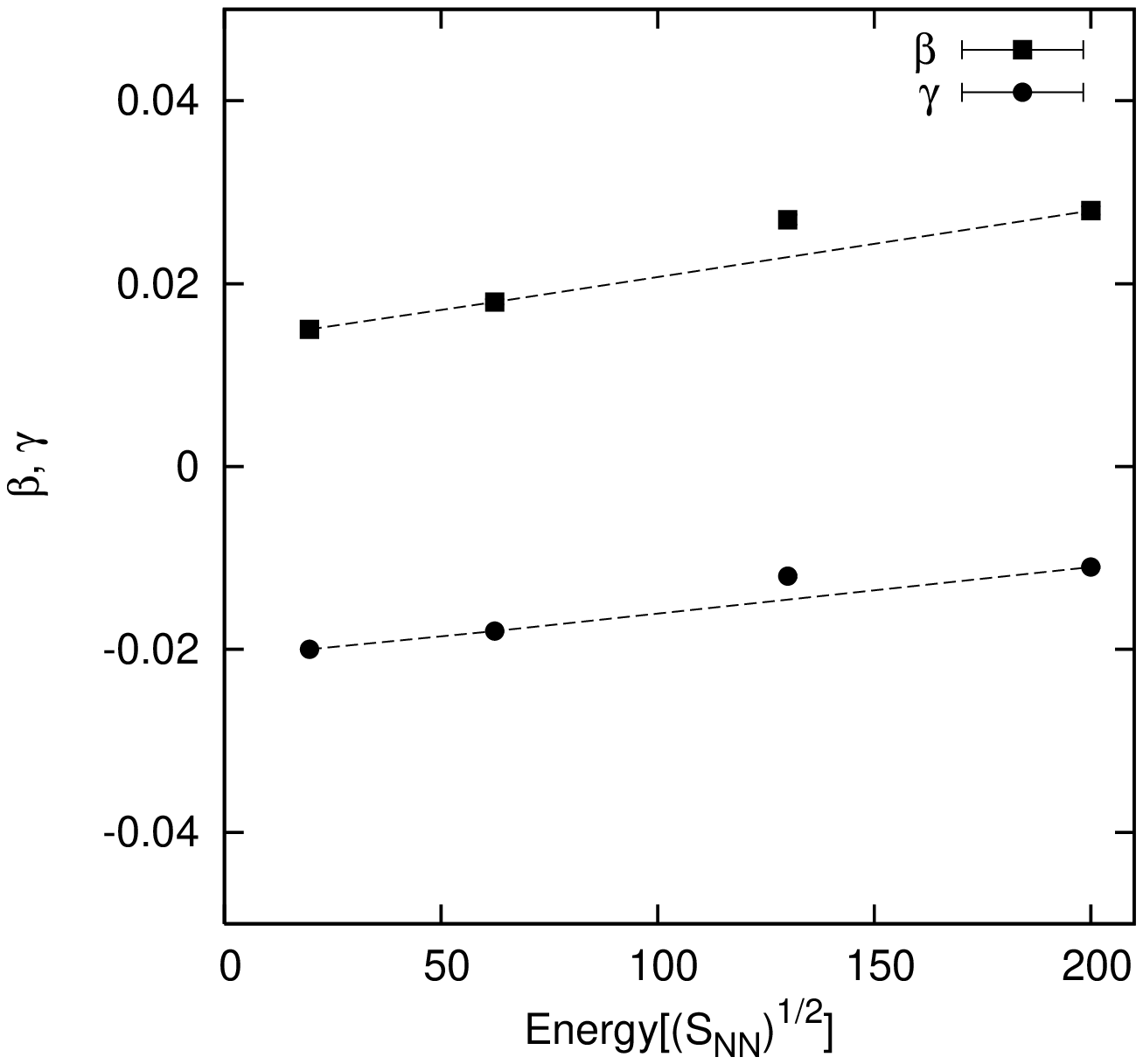}
\caption{Variation of $\beta$ and $\gamma$ with increasing energy.
All values are taken form Table 11.}
\end{figure}

\end{document}